\documentclass[12pt]{article}
\usepackage{graphics}
\usepackage{a4wide}
\usepackage{amssymb,amsmath}
\usepackage{mathtools}
\usepackage[english]{babel}
\usepackage{subfig}
\usepackage{hyperref}
\usepackage{placeins}
\usepackage{color}
\usepackage{enumerate} % for creating lists
\usepackage{siunitx}

\usepackage{caption}
\title{Statistical properties of thermally expandable particles in soft Rayleigh-Benard convection}

\author{Kim M. J. Alards\\ Department of Applied Physics,\\Eindhoven
  University of Technology,\\P.O. Box 513, 5600 MB Eindhoven, The
  Netherlands \and Rudie P. J. Kunnen\\
Department of Applied Physics,\\Eindhoven
  University of Technology,\\P.O. Box 513, 5600 MB Eindhoven, The
  Netherlands 
\and Herman
  J. H. Clercx\\ 
Department of Applied Physics,\\Eindhoven
  University of Technology,\\P.O. Box 513, 5600 MB Eindhoven, The
  Netherlands 
 \and Federico Toschi\\ 
Department of Applied Physics\\
 Department  of Mathematics and Computer Science\\Eindhoven
  University of Technology,\\P.O. Box 513, 5600 MB Eindhoven, The
  Netherlands\\
  Istituto per le Applicazioni del Calcolo,\\ 
Consiglio Nazionale delle
  Ricerche,\\ Via dei Taurini 19, 00185 Rome, Italy.}

\begin{document}
%
%%%%%%%%%%%%%%%%%%%%%%%%% TITLE  %%%%%%%%%%%%%%%%%%%%%%%%%%
\maketitle
%%%%%%%%%%%%%%%%%%%%%%%%% ABSTRACT  %%%%%%%%%%%%%%%%%%%%%%%%%%
%
\abstract{
The dynamics of inertial particles in Rayleigh-B\'{e}nard convection, where both particles and fluid exhibit thermal expansion, is studied using direct numerical simulations (DNS). We consider the effect of particles with a thermal expansion coefficient larger than that of the fluid, causing particles to become lighter than the fluid near the hot bottom plate and heavier than the fluid near the cold top plate. Because of the opposite directions of the net Archimedes' force on particles and fluid, particles deposited at the plate now experience a relative force towards the bulk. The characteristic time for this motion towards the bulk to happen, quantified as the time particles spend inside the thermal boundary layers  (BLs) at the plates, is shown to depend on the thermal response time, $\tau_T$, and the thermal expansion coefficient of particles relative to that of the fluid, $K = \alpha_p / \alpha_f$. In particular, the residence time is constant for small thermal response times, $\tau_T \lesssim 1$, and increasing with $\tau_T$ for larger thermal response times, $\tau_T \gtrsim 1$. Also, the thermal BL residence time is increasing with decreasing $K$. A one-dimensional (1D) model is developed, where particles experience thermal inertia and their motion is purely dependent on the buoyancy force. Although the values do not match one-to-one, this highly simplified 1D model does predict a regime of a constant thermal BL residence time for smaller thermal response times and a regime of increasing residence time with $\tau_T$ for larger response times, thus explaining the trends in the DNS data well. 
}
 
\maketitle
%
%%%%%%%%%%%%%%%%%%%%%%%%% INTRODUCTION  %%%%%%%%%%%%%%%%%%%%%%%%%%
\section{Introduction}

Inertial particles in thermally driven flows are abundant in both nature and technological applications. In nature typical examples are aerosols in the atmospheric boundary layer \cite{Ackerman2004}, the dynamics of droplets in clouds \cite{Shaw2003,Kostinski2005} or plankton in oceanic flows \cite{Reigada2003,Squires1995}, while in technological applications one can think of spray combustion \cite{Faeth1987,Post2002} or solar collectors \cite{Okada1997}. Particles in flows occur in a wide range of densities; while plankton and algae in the ocean have a density close to that of the carrier fluid, droplets in clouds are in general much heavier than the surrounding fluid. When the particle density is different from the fluid density, inertia will cause particle trajectories to deviate from the fluid stream lines, resulting in a non-homogeneous distribution of particles in the flow \cite{Kuerten2016,Calzavarini2008,Douady1991,Saw2008}. When heat transfer between particles and fluid is not instantaneous, also thermal inertia plays a role. Thermal inertia takes into account the time particles need to adjust their internal temperature to that of the surrounding fluid, which is typically referred to as the thermal response time.  \par

The effect of thermal inertia will be visible in the temperature statistics of the particles. The larger the thermal response time of particles, the more the temperature of particles will deviate from the underlying fluid temperature (at the particle position). When considering a dilute suspension (where particles are not expected to influence the fluid flow or temperature) where the size of particles is independent of temperature, thermal inertia will not influence the motion of  the inertial particles. However, when the volume of the particles does depend on the temperature, thermal inertia can drastically change the trajectories of the inertial particles. For example, bubbles in boiling convection will grow in the warmer spots of the flow and shrink in the cooler spots \cite{Lakkaraju2011,Lakkaraju2013,Oresta2009}, affecting their buoyancy and therefore changing the upward and downward motion of these bubbles. This behavior is not restricted to bubbles, but also, \textit{e.g.}, trajectories of (phase changing) oil droplets \cite{chao} or gel-like particles in non-isothermal flows are expected to be influenced by thermal inertia. \par 

Here we conduct a numerical study on the dispersion of thermally and mechanically inertial particles in Rayleigh-B\'{e}nard convection (RBC), a fluid layer heated from below and cooled from above. The typical flow structure in such a RBC set-up is a large-scale circulation (LSC) of rising hot fluid and descending cold fluid \cite{Ahlers2009}. The temperature dependency of the particle size is included as thermal expansion. This method is not restricted to bubbles but can also deal with fluid--fluid systems or gel-like particles in non-isothermal flows. Not only the thermal expansion of particles, but also that of the fluid is taken into account, such that the volume of both particles and fluid increases (linearly) with increasing temperature. In particular, we study particles with an average density equal to that of the fluid and a thermal expansion coefficient  larger than that of the fluid. In this setting, particles become lighter than the fluid near the hot bottom plate and  heavier than the fluid near the cold top plate. This is expected to induce an enhanced upward or downward motion to the particles, respectively, on top of the motion of plumes near the plates. These plumes were shown to be able to transport inertial particles away from the plates, however (without thermal expansion) particles were eventually deposited at the plates again due to the gravitational force \cite{Lavezzo2011}. By including thermal expansion we expect particles to be transported towards the plates by the LSC, be deposited on the plate due to their mechanical inertia, and stay there for some characteristic residence time to then re-suspend due to their enhanced thermal expansion compared to the fluid. \par 

In this way, thermal expansion of particles prevents them from definitively settling at the horizontal plates. In experiments, even a very small mismatch between fluid and particle density leads to particles getting deposited at the top and bottom plates (as, \textit{e.g.}, in \cite{Joshi2016}). This settling of particles will not only reduce the number of particles inside the bulk flow, but could also have significant effects on the heat transfer. The effect of thermally conductive particles with density very close to that of the fluid on the heat transfer in RBC was investigated experimentally by Joshi \textit{et al.} \cite{Joshi2016}. Particles were found to settle at the walls, depleting the RBC bulk flow of particles and forming a porous layer at the plates that eventually would cause a decrease of the heat transfer. In numerical studies particles are often prevented from getting stuck at the plates by neglecting gravity \cite{Arcen2012,Kuerten2011,Wetchagarun2010}, by pointing gravity in the direction parallel to the walls \cite{Lessani2013,Nakhaei2017} or by removing particles from the flow as soon as they reach one of the plates \cite{Lakkaraju2013,Oresta2009,Oresta2013}. Here, the larger thermal expansion coefficient of particles alone ensures that particles eventually move away from the plates again.  \par

The dynamics of thermally inertial particles (without thermal expansion) in RBC has already been studied numerically in the limit of bubbles (light particles) \cite{Lakkaraju2011,Lakkaraju2013,Oresta2009} and in the limit of particles which are heavier than the fluid \cite{Oresta2013}. In these studies a two-way coupling approach is used, \textit{i.e.} the feedback reaction of particles on the fluid velocity and temperature is included in the momentum and energy equations. It was found that these two-way coupled inertial particles significantly affect the heat transfer due to the mismatch between fluid and particle density. However, as a result of this density mismatch particles will get stuck at the horizontal plates. Here we consider particles with a temperature dependent density but with an average density equal to that of the fluid. In this regime of density ratios particles are not expected to significantly influence the heat transfer and flow structures. A one-way coupling treatment is then sufficient. An example of a system where particles have a density very close to that of the fluid, but also a larger thermal expansion coefficient than the fluid, is a configuration of gel-like particles in water; these particles can consist of a rubber coating filled with a mineral or silicon gel \cite{Engelmann2012}. We study how thermal inertia affects the dynamics and the distribution of such particles in RBC. \par 

In the remainder of this paper we first introduce the numerical set-up in sect. \ref{sec:methodsRBC} by explaining both the RBC flow set-up and the modeling of thermally and mechanically inertial particles. In sect. \ref{sec:Results}, we discuss our results, focusing on the distribution and dynamics of these thermally responsive particles in RBC. Results will be presented for a wide range of thermal response times and for different ratios between the thermal expansion coefficient of the particles and of the fluid. In the last sect. \ref{sec:Conclusion} we will summarize and conclude our findings.

%%%%%%%%%%%%%%%%%%%%%%%%% EXPERIMENTAL AND NUMERICAL METHODS  %%%%
%
\section{Numerical methods} \label{sec:methods}
We study RBC, seeded with thermally and mechanically inertial particles, using direct numerical simulations (DNS). Below we will discuss both the numerical model for RBC and the modeling of the inertial particles. 
\subsection{Rayleigh--B\'{e}nard convection} \label{sec:methodsRBC}
In RBC a fluid is heated from below and cooled from above, inducing a buoyancy driven flow. Control parameters for the RBC set-up are the Rayleigh number, $Ra=\alpha_f g \Delta T H^3 / (\kappa \nu)$, and the Prandtl number, $Pr = \nu / \kappa$, with $\alpha_f$ the thermal expansion coefficient of the fluid, $g$ the gravitational acceleration, $\Delta T$ the temperature difference between the plates, $H$ the height of the RBC cell and $\kappa$ and $\nu$ the thermal diffusivity and the kinematic viscosity of the fluid, respectively. The numerical Rayleigh--B\'{e}nard set-up studied here is bounded above and below by horizontal walls and has periodic boundary conditions in the horizontal directions. The governing dimensionless equations are the incompressible Navier--Stokes and energy equations in the Boussinesq approximation:
\begin{align}
	\pmb{\nabla} \cdot \textbf{u}_f &= 0 \label{eq:cont}, \\
	\frac{\partial \textbf{u}_f}{\partial t} + (\textbf{u}_f\cdot \pmb{\nabla}) \textbf{u}_f  &= - \pmb{\nabla} p +  \sqrt{\frac{Pr}{Ra}} \nabla^2 \textbf{u}_f +  T \hat{\textbf{z}}, \label{eq:NS}  \\
	\frac{\partial T_f}{\partial t} + (\textbf{u}_f \cdot \pmb{\nabla}) T_f &= \frac{1}{\sqrt{Pr Ra}} \nabla^2 T_f, \label{eq:energy}
\end{align}
with $\textbf{u}_f$ the fluid velocity vector, $t$ time, $p$ pressure, $T_f$ the fluid temperature and $\hat{\textbf{z}}$ the vertical unit vector. The equations are non-dimensionalized using $H$ for length, $\Delta T$ for temperature and $t_c=H/U$ for time, based on the free-fall velocity $U \equiv \sqrt{g \alpha_f \Delta T H}$. The non-dimensionalization of the temperature is such that the hot bottom plate has a dimensionless temperature of $T(z=0)=1$ and the cold top plate has a dimensionless temperature of $T(z=1)=0$. The equations are solved with no-slip boundary conditions (BCs) and a fixed temperature at the top and bottom plates, while the domain is periodic in the horizontal directions. The domain size is $ 2H \times 2H \times H$ in the $x$-, $y$- and $z$-directions, respectively,  resulting in an aspect ratio of $\Gamma = 2$. This domain is discretized with \num{256 x 256 x 128} grid points and to also ensure at least ten grid points in the thermal and viscous boundary layers (BLs),  grid refinement is used in the vertical direction. The discretization is performed on a staggered grid using a second-order finite-difference scheme and for the integration a third-order Runge--Kutta method is applied. Details of the numerical scheme can be found in \cite{Verzicco2003,vanderPoel2015}.  In this study the Rayleigh and the Prandtl number are fixed as $Ra =2 \cdot 10^7$ and $Pr=6.7$ (corresponding to water). All fluid properties, corresponding to the average fluid temperature $T_m = 0.5$, are reported in detail in table \ref{tab:fluidprops}. 
\begin{table*}
\centering
\setlength{\tabcolsep}{8pt}
\caption{Fluid properties of the the Rayleigh--B\'{e}nard convection flow studied here (at the average fluid temperature). The reported dimensionless properties are: the kinematic viscosity, $\nu = \bar{\nu} / (UH) $, the thermal diffusivity, $\kappa = \bar{\kappa} / (UH)$, the thermal expansion coefficient, $\alpha_f = \bar{\alpha}_f \Delta T$, the energy dissipation, $\epsilon = \bar{\epsilon} H / U^3$, the Kolmogorov length scale, $\eta = \bar{\eta} / H$, the Kolmogorov time scale, $\tau_{\eta} = \bar{\tau}_{\eta} U / H$, the gravitational acceleration, $g = \bar{g} H / U^2$, the Prandtl number, $Pr$, the Rayleigh number, $Ra$, and the Taylor Reynolds number, $Re_{\lambda}$. The dimensional properties, indicated by the bar, are non-dimensionalized using the cell height $H$, the free-fall velocity $U$ and the temperature difference $\Delta T$. The Taylor Reynolds number is defined as $Re_{\lambda} = u'^2 \sqrt{ 15 /( \epsilon \nu) }$, with $u' = (u_x^{rms} + u_y^{rms} + u_z^{rms})/3$ and  $u_{i}^{rms} = \langle \left[ u_i - \langle u_i \rangle \right]^2 \rangle^{1/2}$, where the average is taken over over the full volume and over time.}
\label{tab:fluidprops}
\begin{tabular}{*{10}{l}}
\hline
\hline \\[-0.3cm]
$\nu$ & $\kappa$ & $\alpha_f$ & $\epsilon$ & $\eta$ & $\tau_{\eta}$ & $g$ & $Pr$ & $Ra$ & $Re_{\lambda}$ \\ \hline \\[-0.25cm]
$5.8 \cdot 10^{-4}$ & $8.6 \cdot 10^{-5}$ & $0.0025$ & $1.7 \cdot 10^{-3}$ & $0.019$ & $0.59$  & $400$  & $6.7$ & $2 \cdot 10^7$ & $17$ \\
\hline 
\hline
\end{tabular}
\end{table*}
\subsection{Thermally expanding inertial particles} \label{sec:methodsparticles}
Particles which experience both thermal and mechanical inertia are evolved in the RBC flow. We treat these particles as point particles, a reasonable assumption when the radius of particles, $r_p$, is smaller than the smallest length scale of the flow, $\eta$, the Kolmogorov length scale. Note that in RBC a second length scale is involved related to the temperature field; the Batchelor length $\eta_B$. In the set-up studied here this length scale is smaller than $\eta$, since $\eta_B = \eta / \sqrt{Pr} \approx 0.4 \eta$. To derive the equation for the thermal inertia it is additionally assumed that the thermal conductivity of the particles is much larger than that of the fluid,  such that the Biot number of the particles  is small, $Bi \ll 1$, and temperature gradients within the particles can be neglected \cite{michaelides1994}. The equation for the velocity of one particle, $\textbf{u}_p$, is based on the Maxey-Riley equation \cite{Maxey1983} and for the temperature of that particle, $T_p$, the approach proposed by Michaelides in  \cite{michaelides1994} is used, such that
\begin{align}
\left( 1+\frac{1}{2\beta} \right) \frac{\text{d}\textbf{u}_p}{\text{d}t} = &\frac{1}{\tau_p} \left( \textbf{u}_f(\textbf{x}_p) - \textbf{u}_p \right)(1 + 0.15 Re_p^{0.687}) + \nonumber \\
&\frac{1}{2 \beta} \frac{\text{D} \textbf{u}_f}{\text{D} t} - \left( 1 - \frac{1}{\beta} \right) g \hat{\textbf{z}}, \label{eq:MRparticles} \\[0.4cm]
\frac{\text{d} T_p}{\text{d} t} = &\frac{1}{\tau_T} \left( T_f(\textbf{x}_p) - T_p  \right) (1 + 0.3 Re_p^{1/2}Pr^{1/3}), \label{eq:TEparticles}
\end{align}
where $\textbf{u}_f(\textbf{x}_p)$ and $T_f(\textbf{x}_p)$ are the fluid velocity and the fluid temperature at the position of the particle, $\textbf{x}_p$, respectively. Here $\beta  = \rho_p / \rho_f$ is the ratio between the density of that particle and the fluid density,  $Re_p = 2 r_p  | \textbf{u}_p - \textbf{u}_f(\textbf{x}_p) | / \nu$ is the particle Reynolds number and $\tau_p$ and $\tau_T$ are the viscous and thermal response times, respectively. These are defined as: 
\begin{align}
\tau_p  &= \frac{2 \beta r_p^2}{9 \nu},  \label{eq:taup} \\
\tau_T  &=  \frac{\beta \gamma r_p^2 }{3\kappa}, \label{eq:tauthermal}
\end{align}
where $\gamma = c_p / c_f$ is the ratio between the specific heats of the particle material, $c_p$, and the fluid, $c_f$, \cite{Maxey1983,michaelides1994}. The forces, included on the right-hand side (rhs) of eq. \eqref{eq:MRparticles}, are the Stokes drag, the added mass (also responsible for the pre-factor on the left-hand side) and the gravitational force. In eq. \eqref{eq:TEparticles}, the term on the rhs is analogue to the drag force. Since the particles simulated here have a particle Reynolds number of about $Re_p \sim 10$ it is necessary to include non-linear effects in the drag forces, represented by the factors $(1 + 0.15 Re_p^{0.687})$ in eq. \eqref{eq:MRparticles} \cite{Armenio2001} and $(1 + 0.3 Re_p^{1/2}Pr^{1/3})$ in eq. \eqref{eq:TEparticles} \cite{Ranz1952}. The pressure gradient force and the Basset history force are not included in eq. \eqref{eq:MRparticles}, while these forces might be important in a system where particle and fluid density are similar and $\beta \approx 1$ \cite{Aartrijk2010,Aartrijk2010a}. We verified that ignoring these terms is not influencing the (statistical) measures discussed in this paper and that the most important contributions actually come from the Stokes drag force, added mass force and the buoyancy force. For clarity we therefore choose to not include the Basset history force and the pressure gradient force. In the equation for the thermal inertia  we neglect both the history force and the force analogue to the added mass contribution \cite{michaelides1994}, again after verifying that the contribution of these terms is minor and that the most important contribution comes from the term analogue to the drag force.

As mentioned above eq. \eqref{eq:TEparticles} is valid for $Bi \ll 1$. The Biot number of the particles is defined as $Bi = 2 h_p r_p / k_p$, with $h_p$ and $k_p$ the heat transfer coefficient and the thermal conductivity of particles, respectively. It is possible to express $h_p$ in terms of a particle-Nusselt number, $Nu_p = 2 r_p h_p / k_f$, with $k_f$ the thermal conductivity of the fluid. The Biot number thus becomes $Bi = Nu_p k_f / k_p$, proportional to the ratio between the thermal conductivity of the fluid and the particles. In a suspension of solid particles in a fluid, $k_p \gg k_f$ and the assumption of $Bi \ll 1$ is indeed valid. However, in fluid--fluid systems $Bi \sim \mathcal{O}(1)$ making temperature differences  between the core and the surface of particles  possible. We expect that this will not significantly effect our results and will at most result in an additional delay in the heat transfer  between the particle and the surrounding fluid. This will lead to a larger `effective' thermal response time and since results are presented in a wide range of thermal response times we expect that our results are also applicable to the case of $Bi \sim \mathcal{O}(1)$.  \par 

Particles and fluid both exhibit thermal expansion with a different thermal expansion coefficient, where the thermal expansion coefficient of the particles is chosen to be larger than that of the fluid, such that $\alpha_p > \alpha_f$. The densities of the fluid and the particles are assumed to decrease linearly with the temperature fluctuations of the fluid ($T_f' = T_f - T_m$) and the fluctuations in the particle temperature ($T_p' = T_p - T_m$): 
\begin{align}
\widetilde{\rho}_f &= 1 - \alpha_f T_f', \label{eq:rhof} \\
\widetilde{\rho}_p &= 1-\alpha_p T_p', \label{eq:rhop}
\end{align}
where the densities of the particles and the fluid at the average temperature are set to unity, $\rho_p(T_p=T_m)  = \rho_f(T_f=T_m) = 1$, without loss of generality. Now also the density ratio is temperature dependent, as
\begin{equation}
\widetilde{\beta} = (1-\alpha_p T_p') / (1-\alpha_f T_f'). \label{eq:beta}
\end{equation}
Due to the thermal expansion also the size of the particles depends on the temperature fluctuations. Under the assumption that temperature fluctuations are small (as also assumed by the Boussinesq approximation for eq. \eqref{eq:NS}) and by using a Taylor expansion, the radius of particles follows as
\begin{equation}
\widetilde{r}_p = r_p \left( 1 + \frac{1}{3} \alpha_p T_p' \right), \label{eq:rp}
\end{equation}
where $\widetilde{r}_p$ is the temperature dependent radius, while $r_p$ is the radius of particles at $T_p=T_m$ and where higher order terms have been ignored. \par 

Since the viscous and thermal response times depend on both the density ratio and the particle radius, they have to be updated accordingly such that:
\begin{align}
\widetilde{\tau}_p  &= \frac{ 2 r_p^2 }{9 \nu} \frac{1-\alpha_p T_p'}{1-\alpha_f T_f'} \left( 1 + \frac{1}{3}\alpha_p T_p' \right)^{2} \approx \tau_p \frac{1- \frac{1}{3} \alpha_p T_p'}{1-\alpha_f T_f'}, \label{eq:taupT} \\[0.4cm] 
\widetilde{\tau}_T &= \frac{\gamma r_p^2}{3 \kappa} \frac{1-\alpha_p T_p'}{1-\alpha_f T_f'} \left( 1 + \frac{1}{3} \alpha_p T_p' \right)^{2} \approx \tau_T \frac{1- \frac{1}{3} \alpha_p T_p'}{1-\alpha_f T_f'}, \label{eq:tauthermalT}
\end{align}
where $\tau_p$ and $\tau_T$ are the particle and thermal response times at $T_p=T_f=T_m$, respectively, and we again neglect higher order terms. To complete the implementation of thermal expansion, the parameters  $\beta$, $\tau_p$ and $\tau_T$, in eqs. \eqref{eq:MRparticles} and \eqref{eq:TEparticles} have to be replaced by the temperature dependent variables $\widetilde{\beta}$, $\widetilde{\tau}_p$ and $\widetilde{\tau}_T$, respectively. Also the particle Reynolds number is now based on the temperature dependent radius $\widetilde{r}_p$. \par

The typical time these thermally responsive particles spend at the plate in order to adjust their density enough to escape the BLs, is expected to depend on the ratio between the specific heats of the particle material and the fluid, $\gamma$. Therefore we study particles in a wide range of thermal response times, $\tau_T$. On top of this, we introduce a key parameter for this study, $K = \alpha_p / \alpha_f$, being the ratio between the thermal expansion coefficient of the particle and that of the fluid. Three different values of this parameter $K$ are studied: $K =1.1$, $K = 2$ and $K = 10$, as also reported in table \ref{tab:particleprops}. The applications mentioned in the introduction, gel-like like particles in water and oil-water configurations, would  fall in the range of $1.1 \lesssim K \lesssim 2$. Here, $K = 10$ is added to also study a more extreme case.  For each value of $K$, ten different particle families are included in the simulation; one family consisting of passive traces and nine families of thermally responsive particles, with $0.05 \leq \tau_T \leq 10$ as reported in table \ref{tab:particleprops}. These thermal response times correspond  to a range of $0.13 \leq \gamma \leq 26$. In general $\gamma \sim \mathcal{O}(1)$ for solid--fluid or fluid--fluid systems. To give an estimate of the corresponding thermal response times; in a range of $0.3 \leq \gamma \leq 3$ the thermal response times would be $0.12 \leq \tau_T \leq 1.2$. Here we again add extreme values of both smaller and larger $\gamma$ to understand how the systems converges in the limit of very small and very large thermal response times. In this parameter range the density ratio varies between $0.96 <  \widetilde{\beta} <  1.04$. Within this range of density ratios particles are not expected to influence the flow structures and the heat transfer and therefore a one-way coupling approach is sufficient. In total nine different particle families are simulated for 300 dimensionless time units, where the number of particles in each family is $1.6 \cdot 10^5$.  A detailed overview of the particle properties is given in table \ref{tab:particleprops}. \par  
\begin{table}
\centering
\caption{Particle properties of the thermally responsive particles (TRP), simulated in Rayleigh-B\'{e}nard convection. Three different simulations are performed with tracers (family 0) and thermally responsive particles (families 1--9), for three different ratios between the thermal expansion coefficient of the particles and that of the fluid, $K = \alpha_p / \alpha_f$. Here $r_p$, $\tau_p$ and $\beta$ are the particle radius, the drag response time and the ratio between the particle and fluid density at the mean temperature $T_m$, respectively. The properties of the different particle families  at the average particle and fluid temperature, are reported at the bottom of the table, where $\gamma = c_p / c_f$, with $c_p$ and $c_f$ the specific heat of the particles and the fluid, respectively, and $\tau_T$ is the thermal response time.} 
\label{tab:particleprops}
\begin{tabular}{ *{5}{p{1.2cm}}  }
\hline
\hline
$K $  & $r_p$ & $\alpha_p$ & $\tau_p$  & $\beta$   \\
\hline
$1.1$  & $0.01$ & $0.00275$ & $ 0.038$  & $1$ \\
$2$  & $0.01$ & $0.005$ & $ 0.038$ & $1$   \\
$10$  & $0.01$ & $0.025$ & $ 0.038$ & $1$ \\
\hline
\hline
\multicolumn{2}{l}{particle family} & $\gamma$  & $\tau_T$  & type \\
\hline
0 & & - & - & tracer \\
1 & & 0.13 & 0.05 & TRP \\ 
2 & & 0.26 & 0.1& TRP \\
3 & & 0.65 & 0.25 & TRP \\
4 & & 1.3 & 0.5 & TRP \\
5 & & 2.6 & 1 & TRP \\
6 & & 5.2 & 2 & TRP \\
7 & & 10 & 4 & TRP  \\
8 & & 16 & 6 & TRP \\
9 & & 26 & 10 & TRP   \\
\hline 
\hline
\end{tabular}
\end{table}
%
%%%%%%%%%%%%%%%%%%%%%%%%% RESULTS  %%%%%%%%%%%%%%%%%%%%%%%%%%%%%%%%
\section{Results} \label{sec:Results}
\subsection{Spatial distribution of thermally expandable particles} \label{sec:ResultsDistribution}

We investigate the dynamics of thermally responsive inertial particles in Rayleigh--B\'{e}nard convection, where we include thermal expansion of both particles and of fluid. In particular, the thermal expansion coefficient of particles is larger than that of the fluid, so that particles react to the temperature fluctuations stronger. Since in RBC the temperature gradients are largest in the thermal BLs while the temperature in the bulk fluctuates around the average temperature \cite{Kerr1996,Ahlers2009}, particles are expected to distribute differently in the bulk than in the thermal BLs when thermal expansion is included. To study the vertical distribution of particles, we compute the particle number density, $n_i$, as a function of $z$. First the RBC cell is subdivided into 250 horizontal slabs of size $\Delta z = 0.004 H$, with central vertical position $z_i$. The number density in each slab is computed as the time averaged number of particles in the slab divided by the slab volume; $\langle N_i \rangle / V_i$, where $V_i = \Delta_z L_x L_y$. Finally this number density is normalized by the total number density $N_{tot} / V_{tot}$, where $N_{tot} = 1.6 \cdot 10^5$ (for each particle family) and $V_{tot} = H L_x L_y$.  In summary this means $n_i = \frac{\langle N_i \rangle}{V_i} / \frac{N_{tot}}{V_{tot}} $. \par

\begin{figure}
\subfloat[$K = 1.1$]{\includegraphics[width=0.5\textwidth]{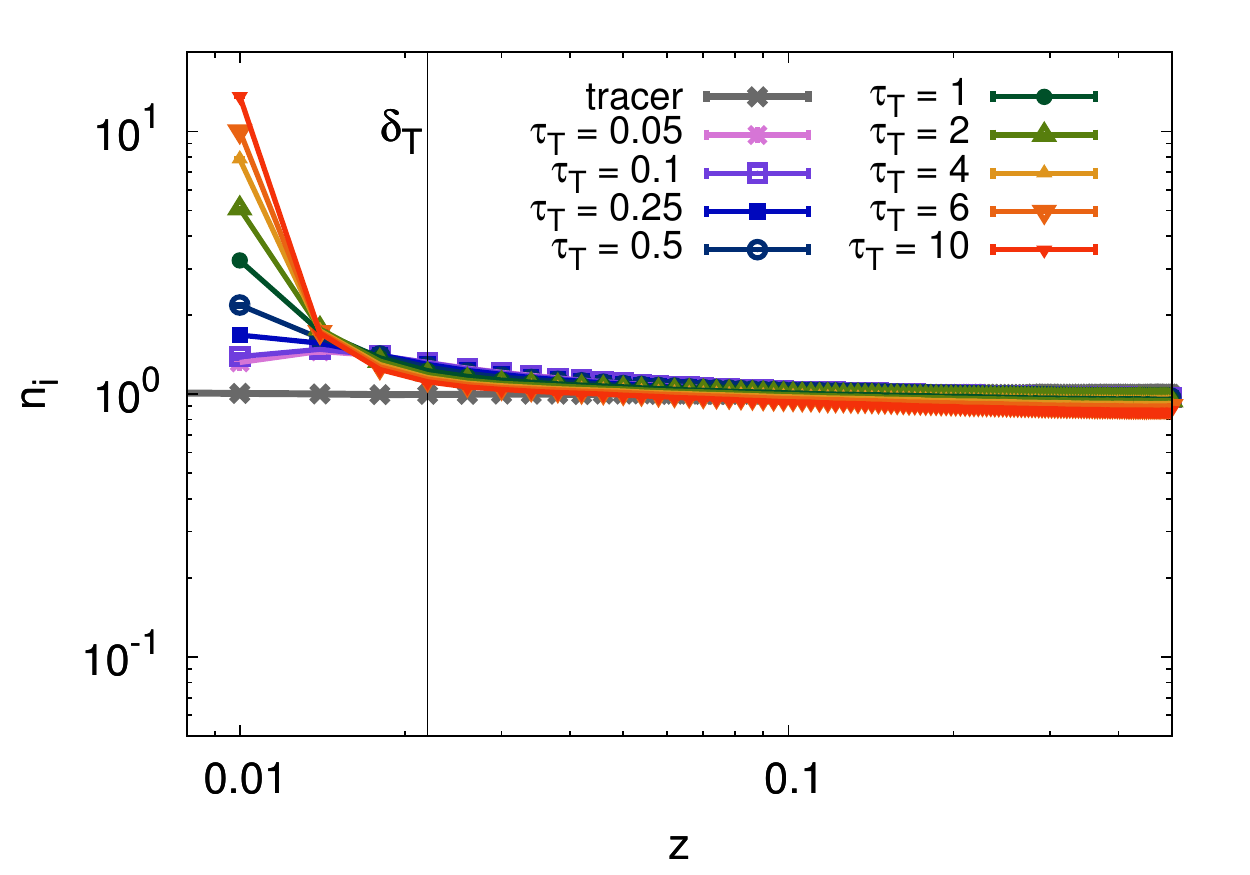} \label{fig:densrun1}}\\
\subfloat[$K = 2$]{\includegraphics[width=0.5\textwidth]{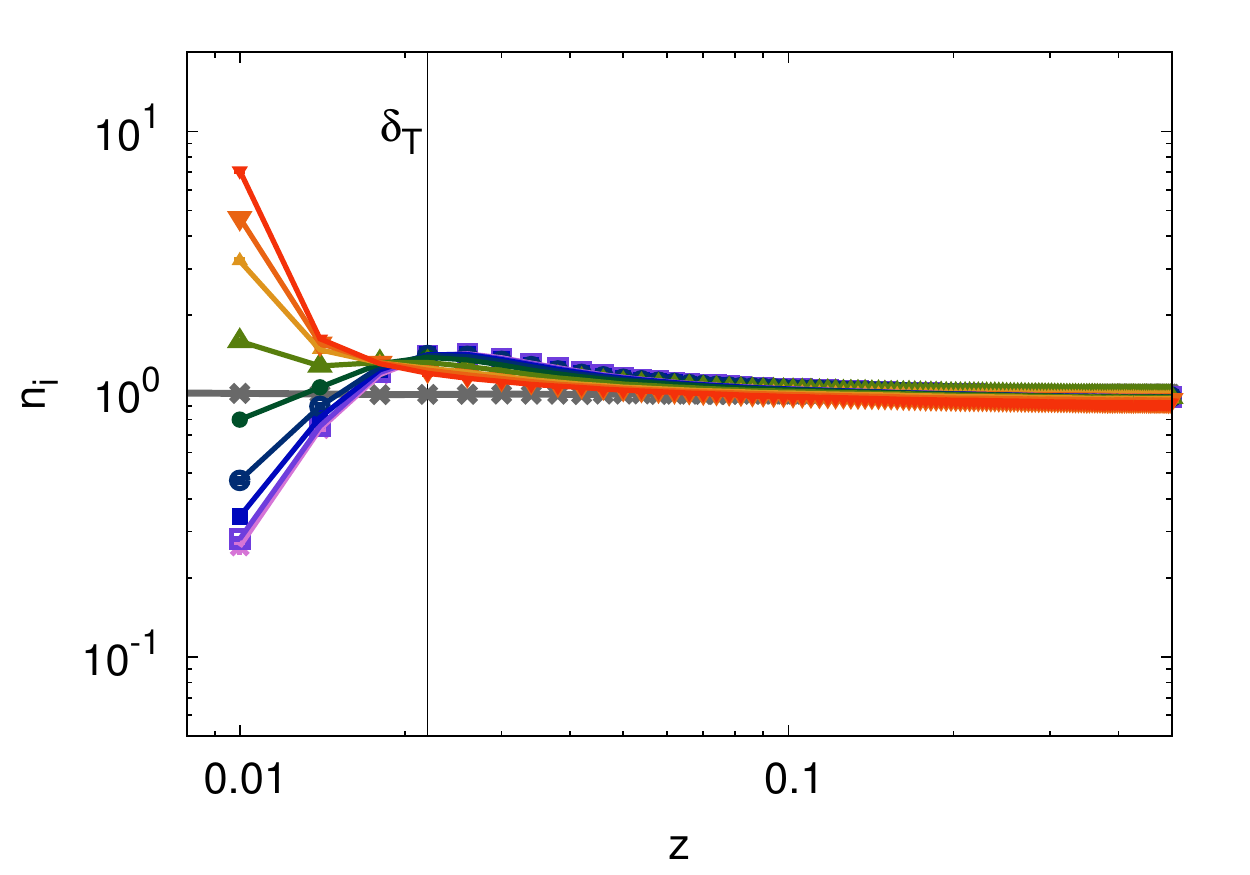} \label{fig:densrun2}}\\
\subfloat[$K = 10$]{\includegraphics[width=0.5\textwidth]{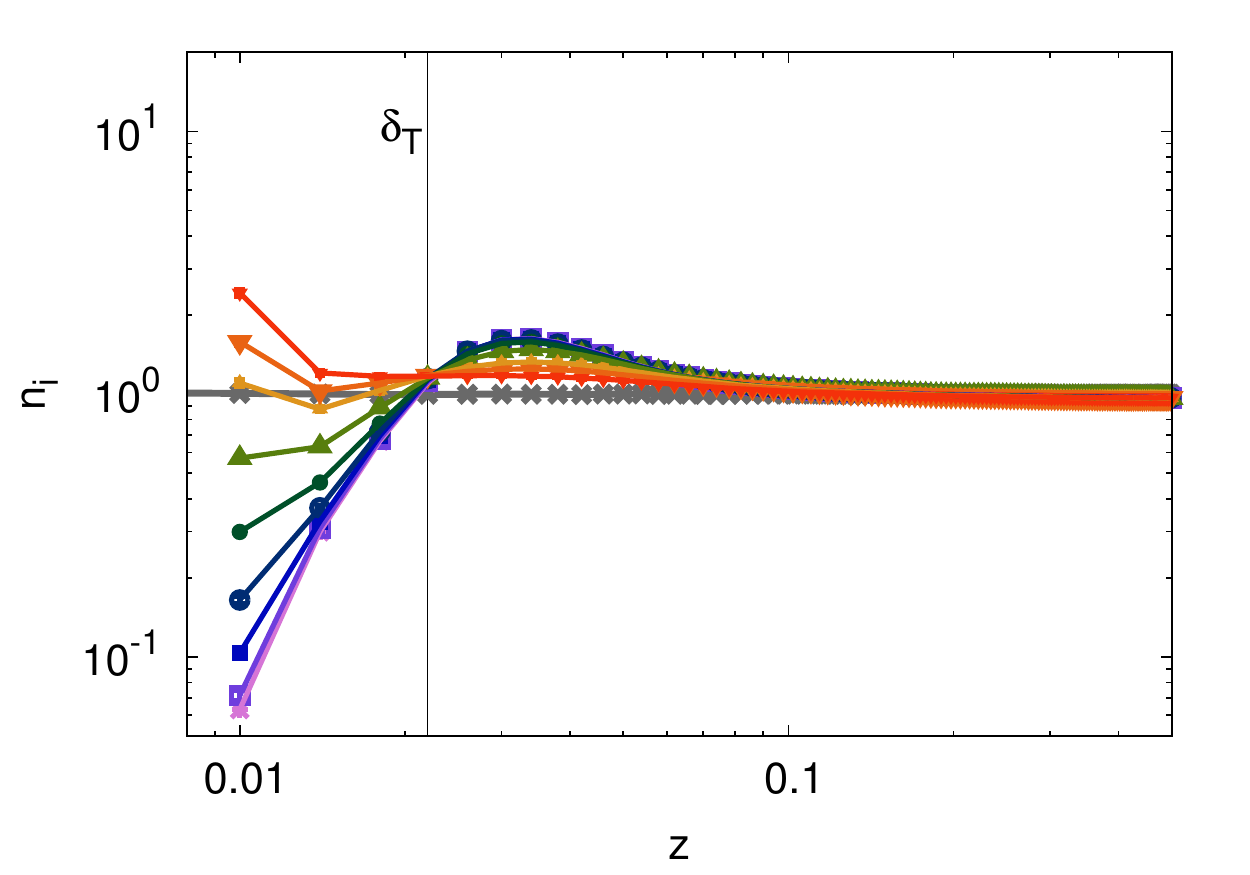} \label{fig:densrun3}}
\caption{ The vertical distribution, $n_i$, of tracers (gray crosses) and thermally responsive particles (different colors) in the Rayleigh--B\'{e}nard cell. Results are shown for three different values of $K$: (a) $K = 1.1$,  (b) $K = 2$ and (c) $K = 10$ and for different $\tau_T$ as reported in the legend of panel (a) (see also table \ref{tab:particleprops}). The solid vertical line shows the thermal boundary layer thickness, $\delta_{T} = 0.022 H$. Because of symmetry we only show the lower half of the domain, where $0<z<0.5 H$. Error bars are estimated as the deviation from this symmetry and are falling within the symbol size. }
\label{fig:dens}
\end{figure}

In fig. \ref{fig:dens}, we show $n_i$ for the three different values of $K$: $K=1.1$, $K=2$ and $K=10$ and different values of $\tau_T$ between $\tau_T = 0.05$ and $\tau_T = 10$. As a reference the distribution of fluid tracers is also shown with gray lines with crosses. As expected, fluid tracers are distributed uniformly such that $n_i = 1$. Note that these fluid tracers have no thermal and mechanical inertia ($\tau_T = \tau_p = 0$) and that they are therefore not affected by thermal expansion. The thermal BL thickness, $\delta_T = 0.022H$, is computed as the position of the maximum root-mean-square temperature and is indicated in fig. \ref{fig:dens} by the vertical black lines. First, we observe that the number of particles inside the thermal BL is increasing with increasing thermal response times compared to the uniform distribution $n_i  = 1$. Particles with a larger thermal response time need more time to heat up  (cool down) at the bottom (top) plate, hence there will be more particles close to the plates on average. Furthermore, when comparing the three different panels, it is observed that this number of particles at the plate is larger for lower values of $K$. Particles with a larger thermal expansion coefficient compared to that of the fluid react very strongly to temperature fluctuations and even a small temperature change can lead to a huge change in their mass density. Consequently, particles move away from the plates faster and the number of particles at the plates decreases. For $K = 2$ and $K = 10$ we observe a regime where $n_i < 1$ for $\tau_T \lesssim 2$ and $\tau_T \lesssim 4$, respectively. Here particles escape the BLs so fast that there is a depletion of particles in the thermal BLs, compared to the average distribution $n_i = 1$. A depletion in the BLs results in an increase of particles in the bulk, indicated by the peaks in fig.s \ref{fig:densrun2} and \ref{fig:densrun3} for $z \gtrsim \delta_T$, which become more prominent for larger $K$ and for smaller $\tau_T$.  \par
\begin{figure*}
\centering
\vspace{-3.5cm} %reduce to -1cm without preprint
\subfloat[temperature field]{\includegraphics[height=6.25cm]{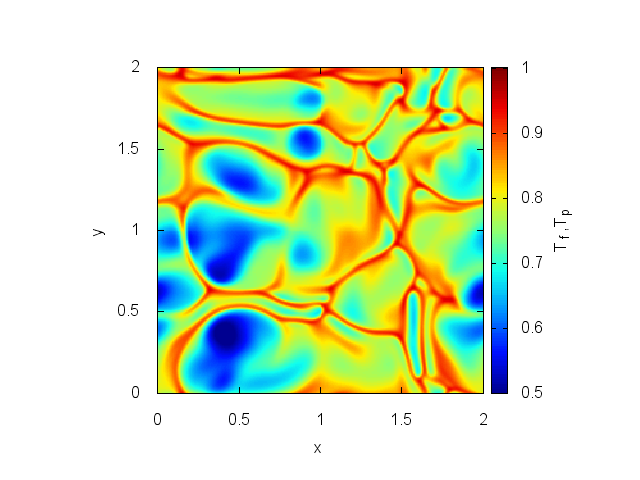} \label{fig:Tfield} } \\
\subfloat[$K = 1.1, \tau_T = 0.1$]{\includegraphics[trim=0 2.0cm 0 2.0cm, clip, height=4.8cm]{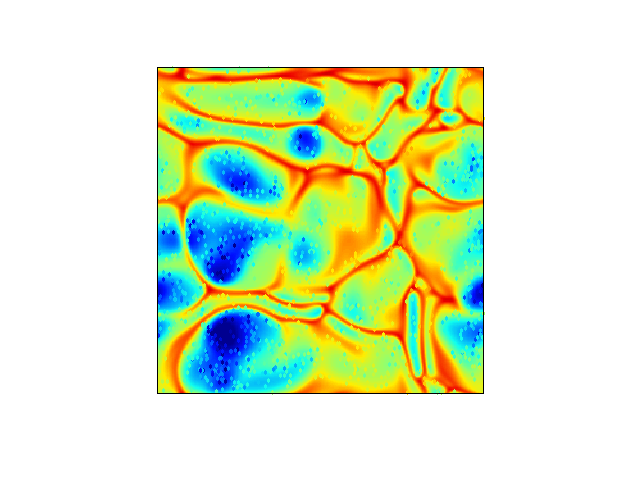} \label{fig:fieldrun1fam2} }%\hspace{-2cm}
\subfloat[$K = 1.1, \tau_T = 4$]{\includegraphics[trim=0 2.0cm 0 2.0cm, clip, height=4.8cm]{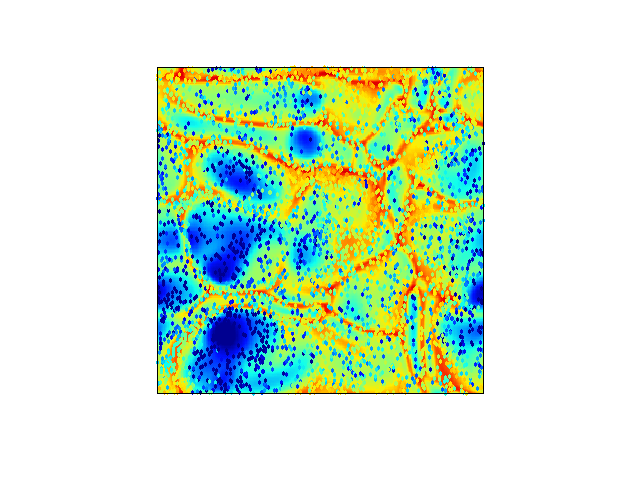} \label{fig:fieldrun1fam7} } \\
\subfloat[$K = 2, \tau_T = 0.1$]{\includegraphics[trim=0 2.0cm 0 2.0cm, clip, height=4.8cm]{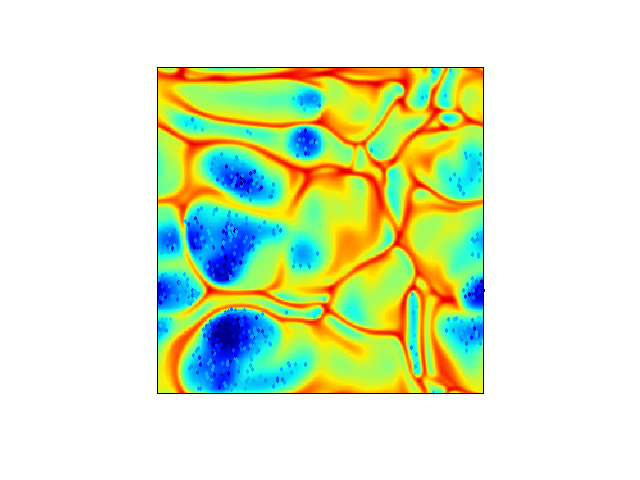} \label{fig:fieldrun2fam2} }%\hspace{-2.0cm}
\subfloat[$K = 2, \tau_T = 4$]{\includegraphics[trim=0 2.0cm 0 2.0cm, clip, height=4.8cm]{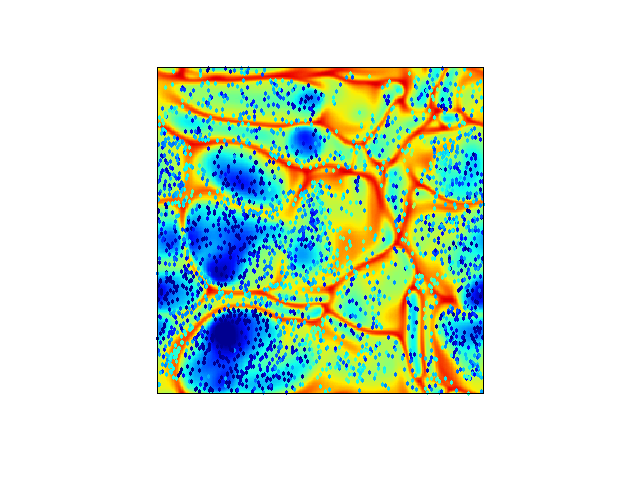} \label{fig:fieldrun2fam7} } \\
\subfloat[$K = 10, \tau_T = 0.1$]{\includegraphics[trim=0 2.0cm 0 2.0cm, clip, height=4.8cm]{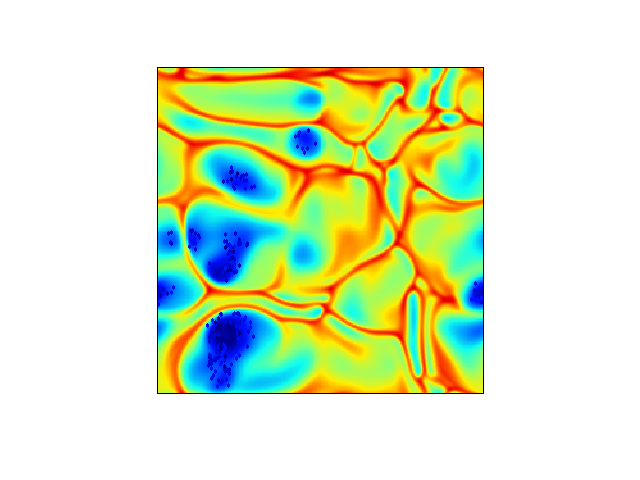} \label{fig:fieldrun3fam2} }%\hspace{-2.0cm}
\subfloat[$K = 10, \tau_T = 4$]{\includegraphics[trim=0 2.0cm 0 2.0cm, clip, height=4.8cm]{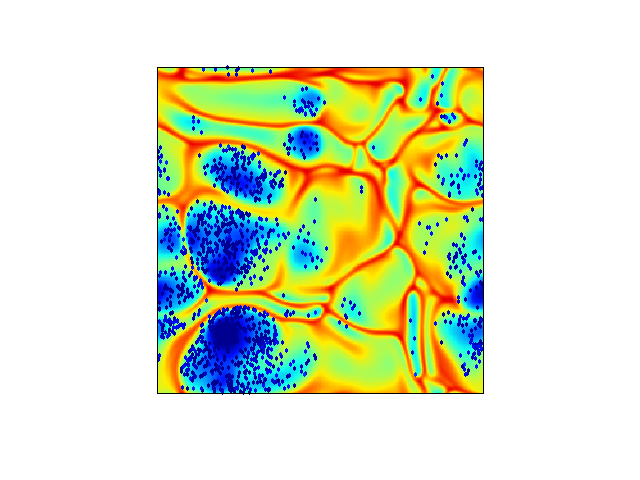} \label{fig:fieldrun3fam7} } \\
\vspace{0.2cm}
\caption{(a) The fluid temperature, $T_f$, in a horizontal plane at $z=0.012 H$ in the Rayleigh--B\'{e}nard cell. (b-g) The same temperature field, together with particles with vertical position $z_p < 0.015 H$ for different values of $K$ and $\tau_T$: (b) $K = 1.1, \tau_T = 0.1$, (c) $K = 1.1, \tau_T = 4$, (d) $K = 2, \tau_T = 0.1$, (e) $K = 2, \tau_T = 4$, (f) $K = 10, \tau_T = 0.1$ and (g) $K = 10, \tau_T = 4$.  Axes and colorbars are as in panel (a) and the color of the particles encodes their temperature, $T_p$. }
\label{fig:densfield}
\end{figure*}

The particle number density, as shown in fig. \ref{fig:dens}, is an average quantity and does not give information on the particle distribution in the horizontal directions. To understand how particles distribute horizontally with respect to the typical temperature profiles in the RBC cell, we visualize the temperature field at $z= 0.012 H$ without particles in fig. \ref{fig:Tfield} and with different types of thermally responsive  particles with vertical position $ z_p < 0.015 H$ in figs. \ref{fig:fieldrun1fam2} - \ref{fig:fieldrun3fam7}, where particles are colored by their temperature. For each value of $K$ ($K = 1.1, K = 2$ and $K = 10$), a situation with a low thermal response time of $\tau_T = 0.1$ and a situation with a large thermal response time of $\tau_T=4$ are shown. First, when focusing on the effect of the thermal response time, it is observed that there are more particles at the plate for larger thermal response times (as already discussed above) and that particles with a lower thermal response time are only found in the colder spots. These particles have a temperature very close to that of the fluid and it is expected that colder heavier particles stay at the plates longer, explaining  why in this regime colder particles are found, clustered in the colder spots of the fluid at the bottom plate in panels (b), (d) and (f). For larger thermal response times, particle and fluid temperature are less correlated and, especially for lower values of $K$, particles are less restricted to the colder areas of the fluid, see panels (c), (e) and (g).   \par
\subsection{Temperature statistics} \label{sec:ResultsTempStats}
\begin{figure*}
\subfloat[$K = 1.1$, average profile]{\includegraphics[width = 0.5\textwidth]{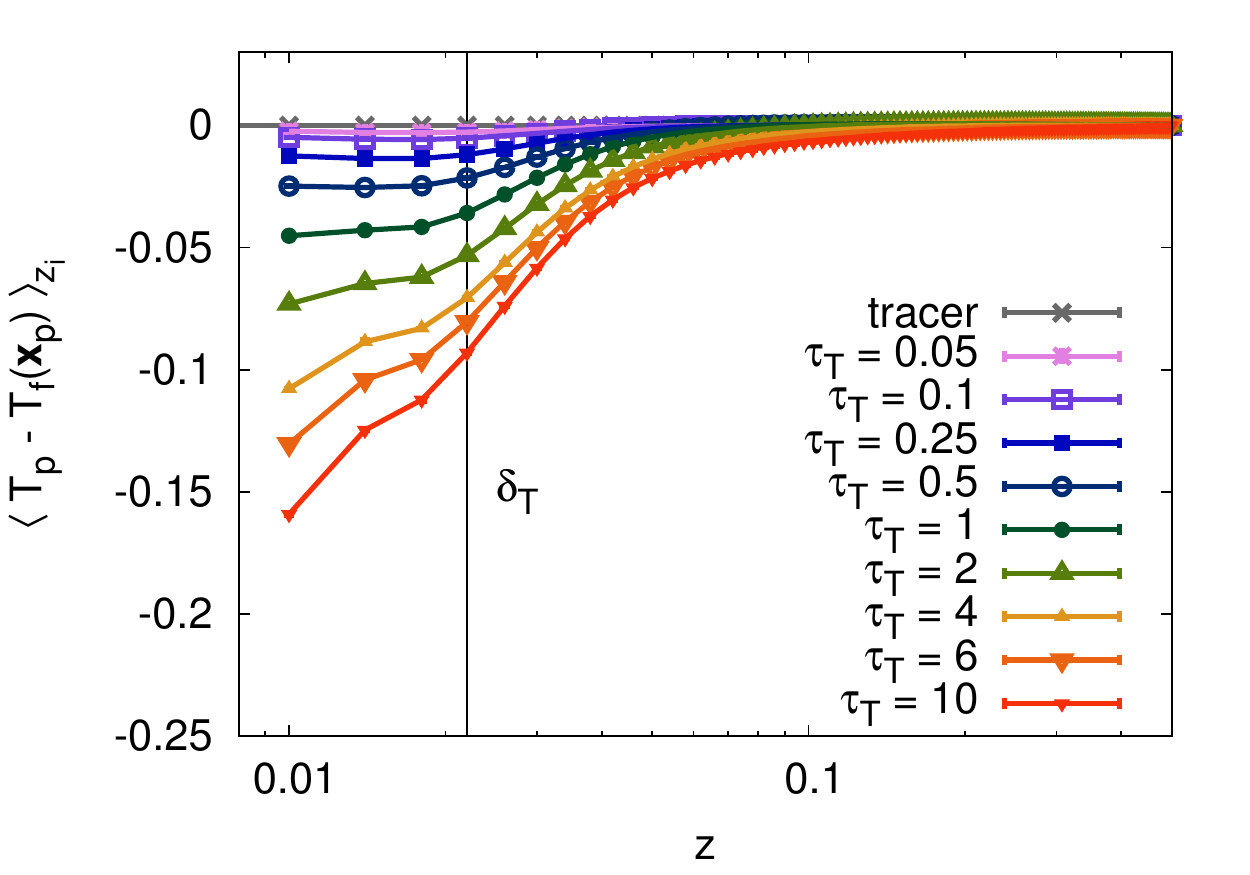} \label{fig:Taveragerun1}}
\subfloat[$K = 1.1$, PDF]{\includegraphics[width=0.5\textwidth]{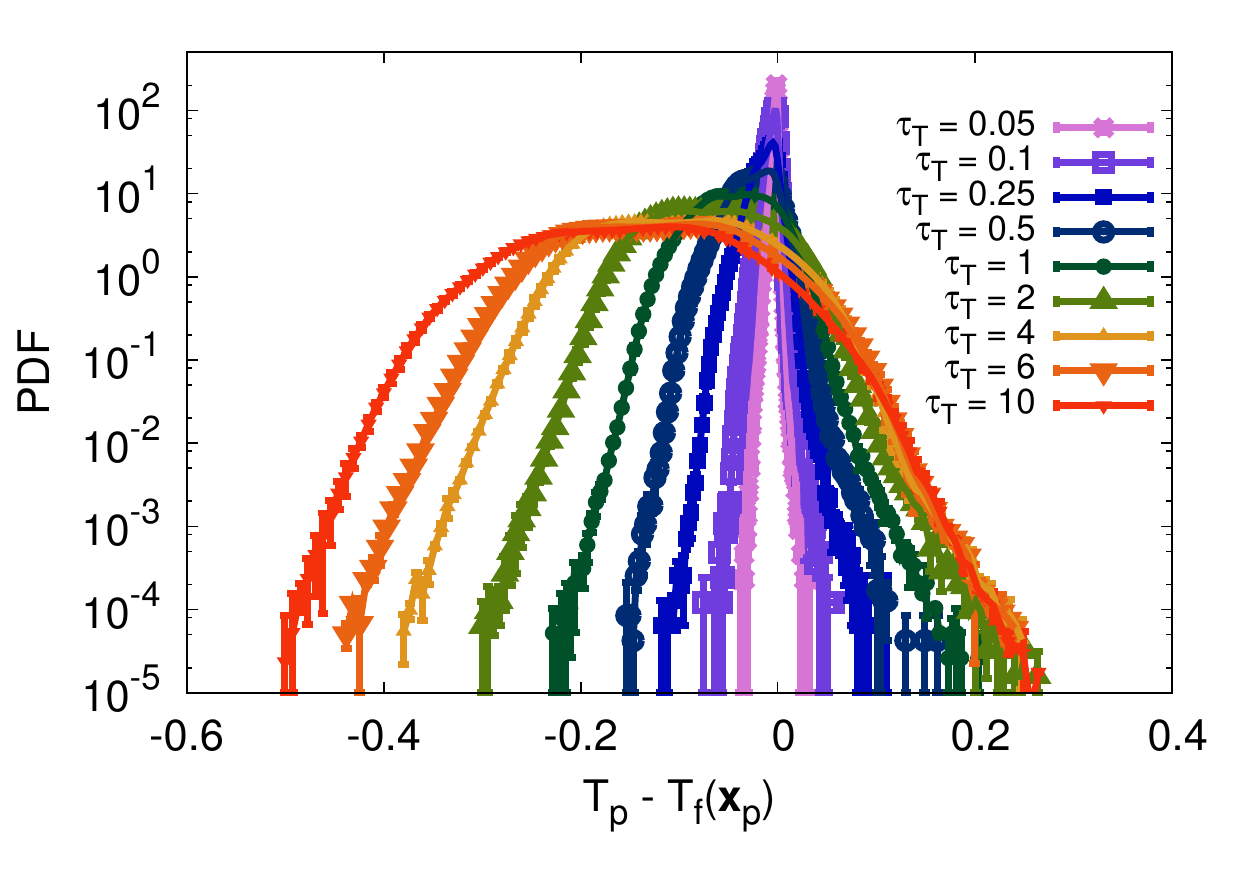} \label{fig:TPDFrun1}}\\

\subfloat[$K = 2$, average profile]{\includegraphics[width = 0.5\textwidth]{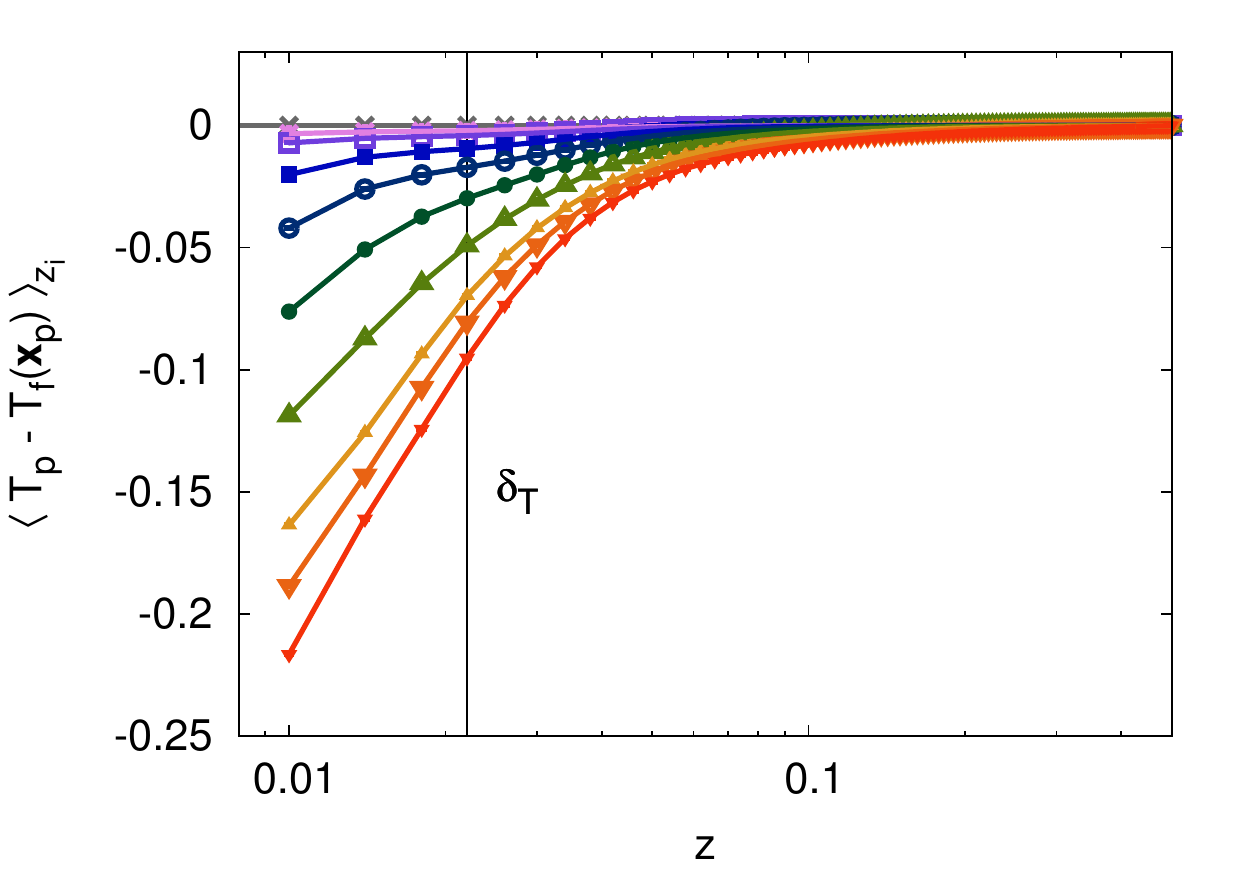} \label{fig:Taveragerun2}}
\subfloat[$K = 2$, PDF]{\includegraphics[width=0.5\textwidth]{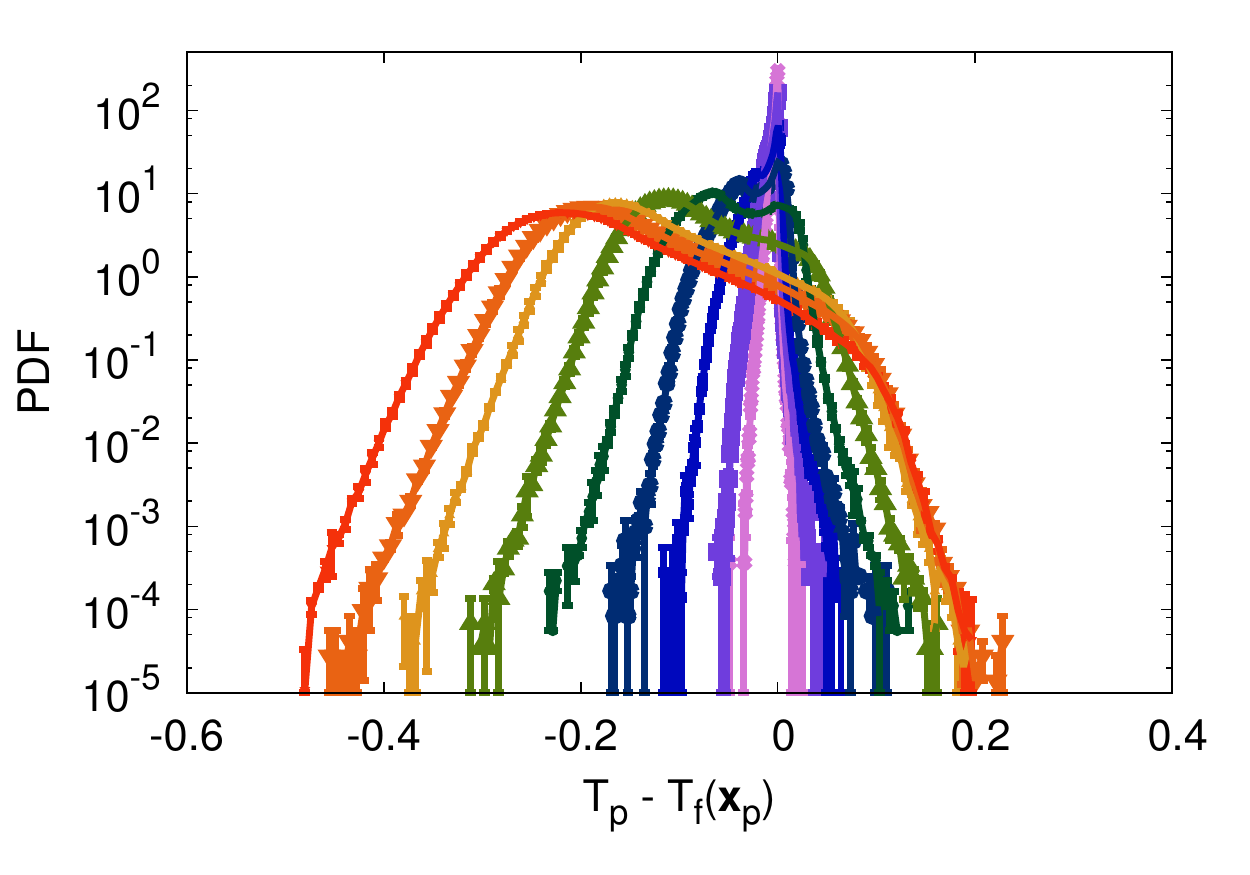} \label{fig:TPDFrun2}}\\ 

\subfloat[$K = 10$, average profile]{\includegraphics[width = 0.5\textwidth]{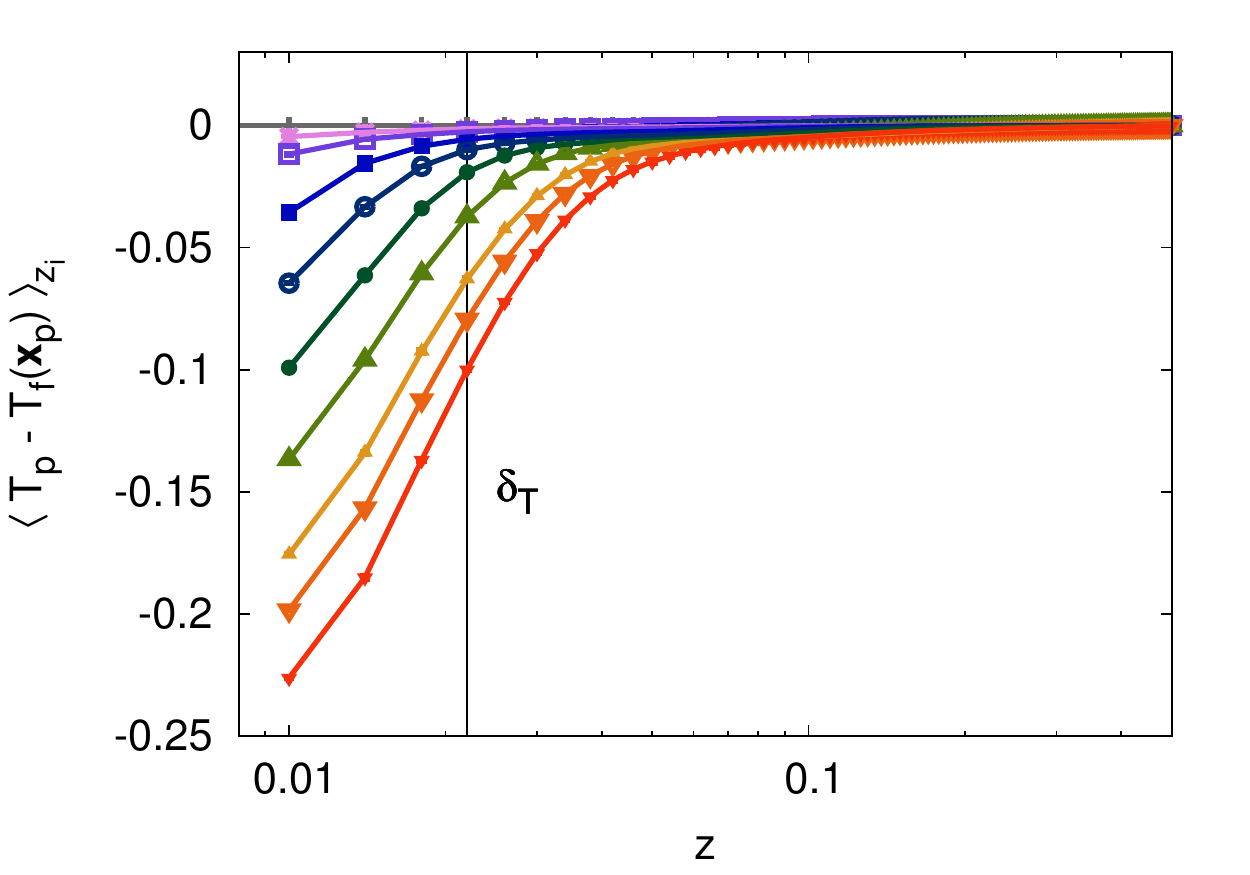} \label{fig:Taveragerun3}}
\subfloat[$K = 10$, PDF]{\includegraphics[width=0.5\textwidth]{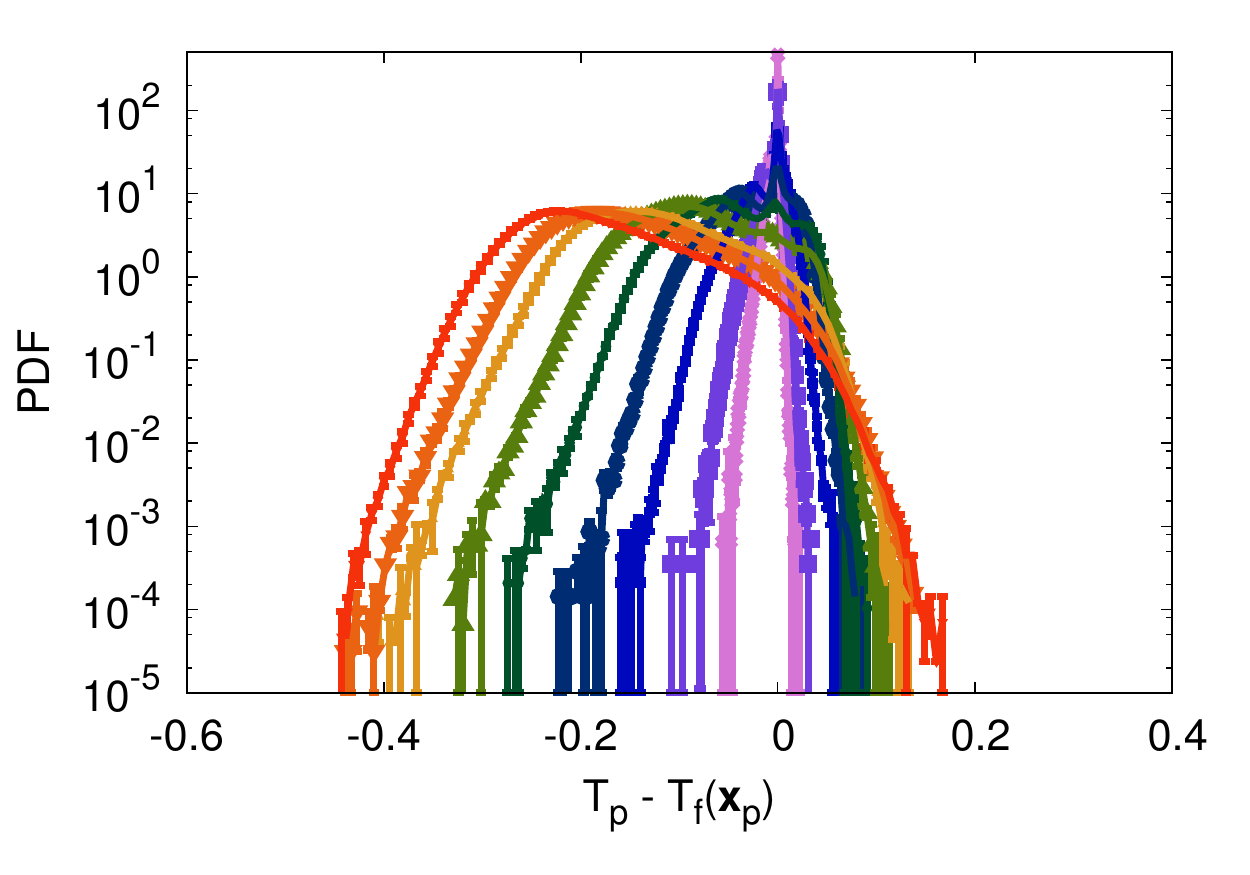} \label{fig:TPDFrun3}}

\caption{(a) The temperature difference $\langle T_p - T_f (\textbf{x}_p) \rangle_{z_i}$, averaged horizontally and in time within horizontal slabs at central vertical position $z_i$ for fluid tracers (gray lines with crosses) and thermally responsive particles (different colors) in Rayleigh--B\'{e}nard convection for $K = 1.1$. (b) PDFs of $T_p - T_f(\textbf{x}_p)$, measured in the thermal boundary layer (BL) at the bottom plate for $K = 1.1$. (c) and (d) show similar results as (a) and (b) but for $K = 2$. (e) and (f) show similar results as (a) and (b) but for $K = 10$. Because of symmetry we only show the first half of the domain in panels (a), (c) and (e) and only results obtained in the BL at the bottom plate in panel (b), (d) and (f), where the thermal BL thickness is $\delta_T=0.022H$. Error bars are estimated as the deviation from this symmetry. In the left panels the error bars fall within the symbol size.  }
\label{fig:difftemp}
\end{figure*}

From fig. \ref{fig:densfield} we expect that the distribution of particles is related to the temperature of the particles, relative to the temperature of the surrounding fluid. This temperature difference is quantified as $T_p - T_f(\textbf{x}_p)$. The average profile of $\langle T_p - T_f(\textbf{x}_p) \rangle_{z_i}$ is computed in vertical slabs of size $\Delta z = 0.004H$ with central vertical position $z_i$ and probability density functions (PDFs) of this quantity are constructed in the BL at the bottom plate ($z_p < \delta_T$). From the left panels of fig. \ref{fig:difftemp} we observe that for all values of $K$ the difference between the (average) particle temperature and the (average) fluid temperature is indeed increasing with increasing $\tau_T$ at the bottom plate.  The PDFs clearly become wider for larger $\tau_T$, again confirming that more extreme temperature differences are found for larger thermal response times as expected. Furthermore, there is an enhanced probability on larger deviations $| T_p - T_f(\textbf{x}_p) |$ for larger values of $\tau_T$, when focusing on the left-hand side (lhs) of the PDFs. The temperature difference of the particles with respect to the fluid at $\textbf{x}_p$ near the bottom plate is also slightly increasing with increasing $K$ as evident when comparing the top, central and bottom panels of fig. \ref{fig:difftemp}. A peak develops on the lhs of the PDFs for increasing $K$, suggesting that there is indeed a larger probability of larger absolute temperature differences for larger values of $K$. This is a result of particles with a large thermal expansion coefficient escaping the warm bottom plate region already for a slight temperature increase. Now, only particles that have a much lower temperature with respect to the fluid temperature stay at the plates longer, resulting in a larger absolute temperature difference $T_p - T_f(\textbf{x}_p)$.  \par 
\subsection{Thermal boundary layer residence time}
\begin{figure}
%\hspace*{-1.3cm}     
\subfloat[$ K = 1.1$]{\includegraphics[width=0.5\textwidth]{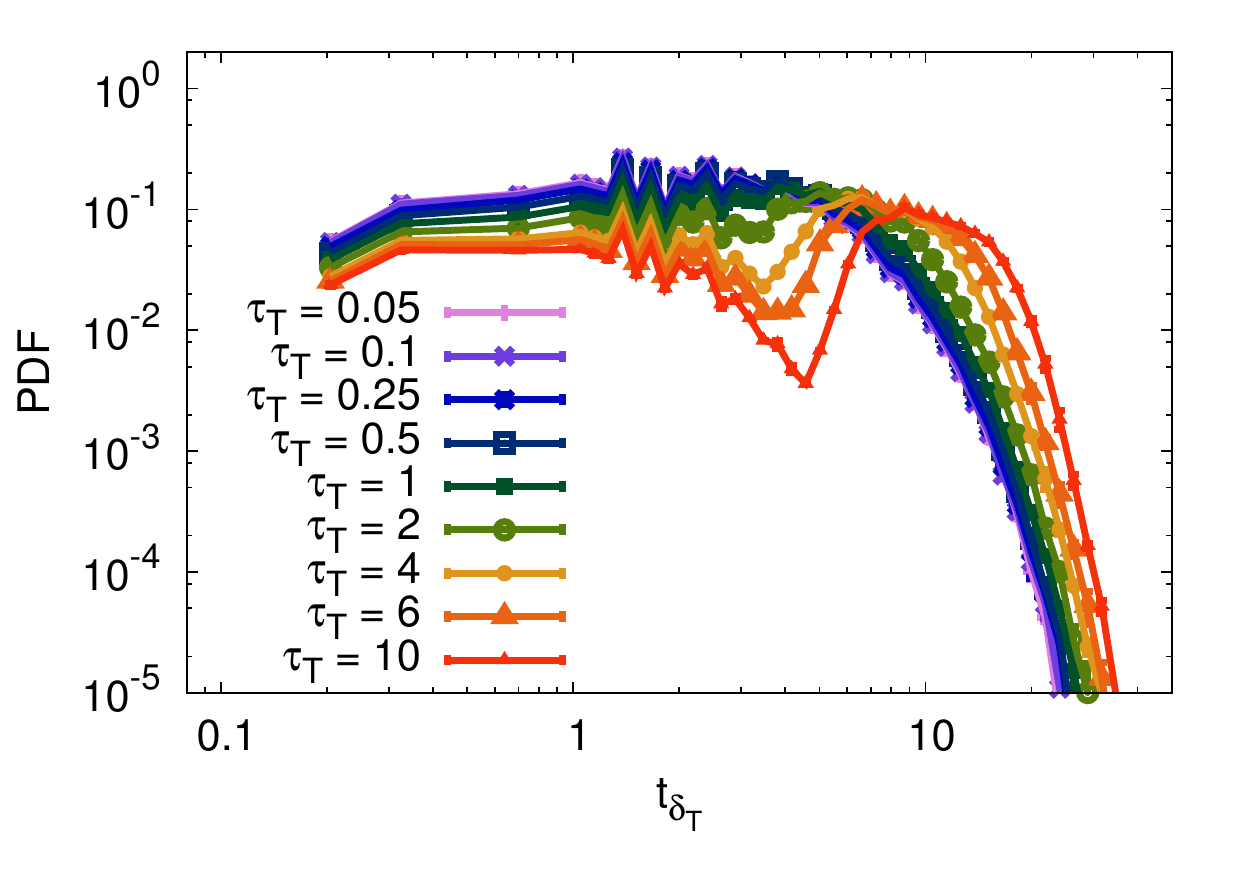} \label{fig:timepdfrun1} }\\
\subfloat[$K = 2$]{\includegraphics[width=0.5\textwidth]{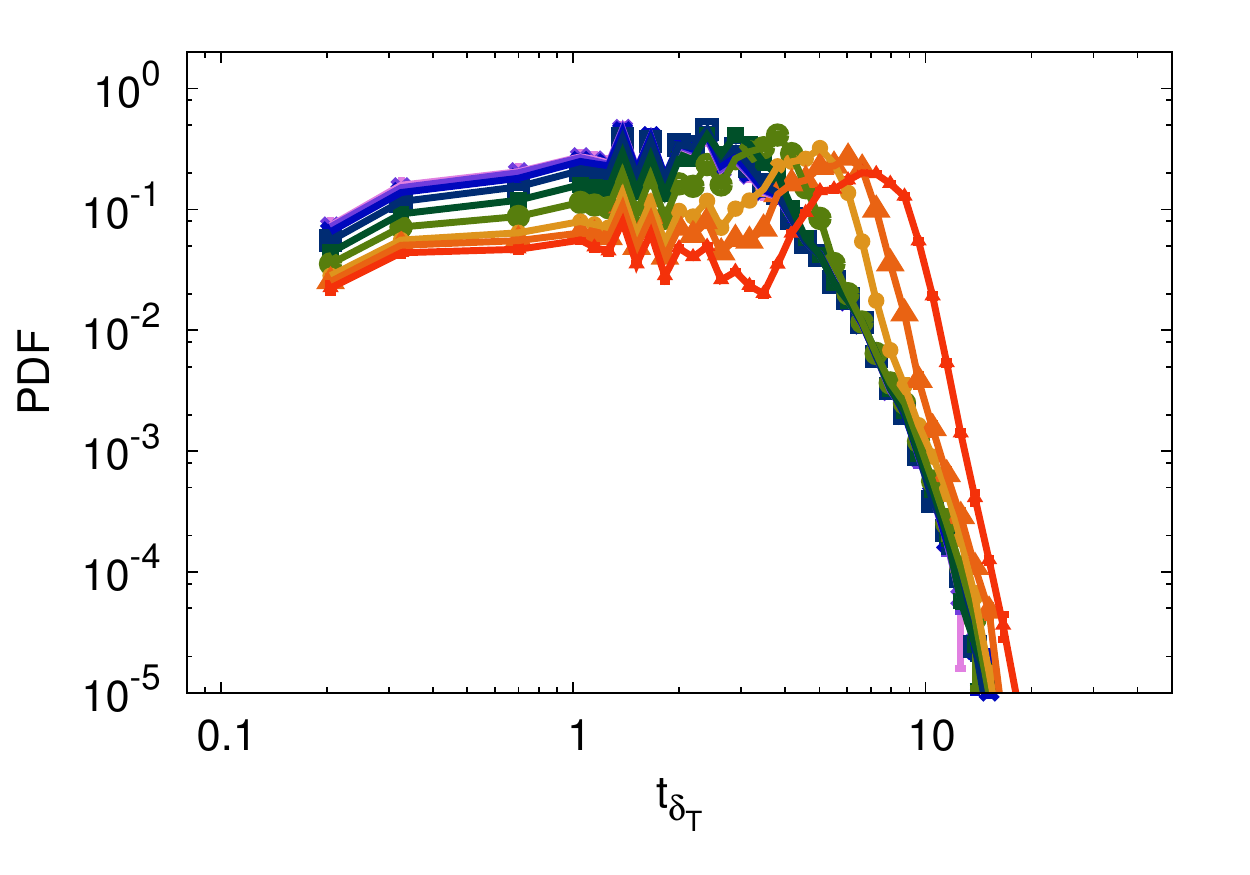} \label{fig:timepdfrun2} }\\
\subfloat[$K = 10$]{\includegraphics[width=0.5\textwidth]{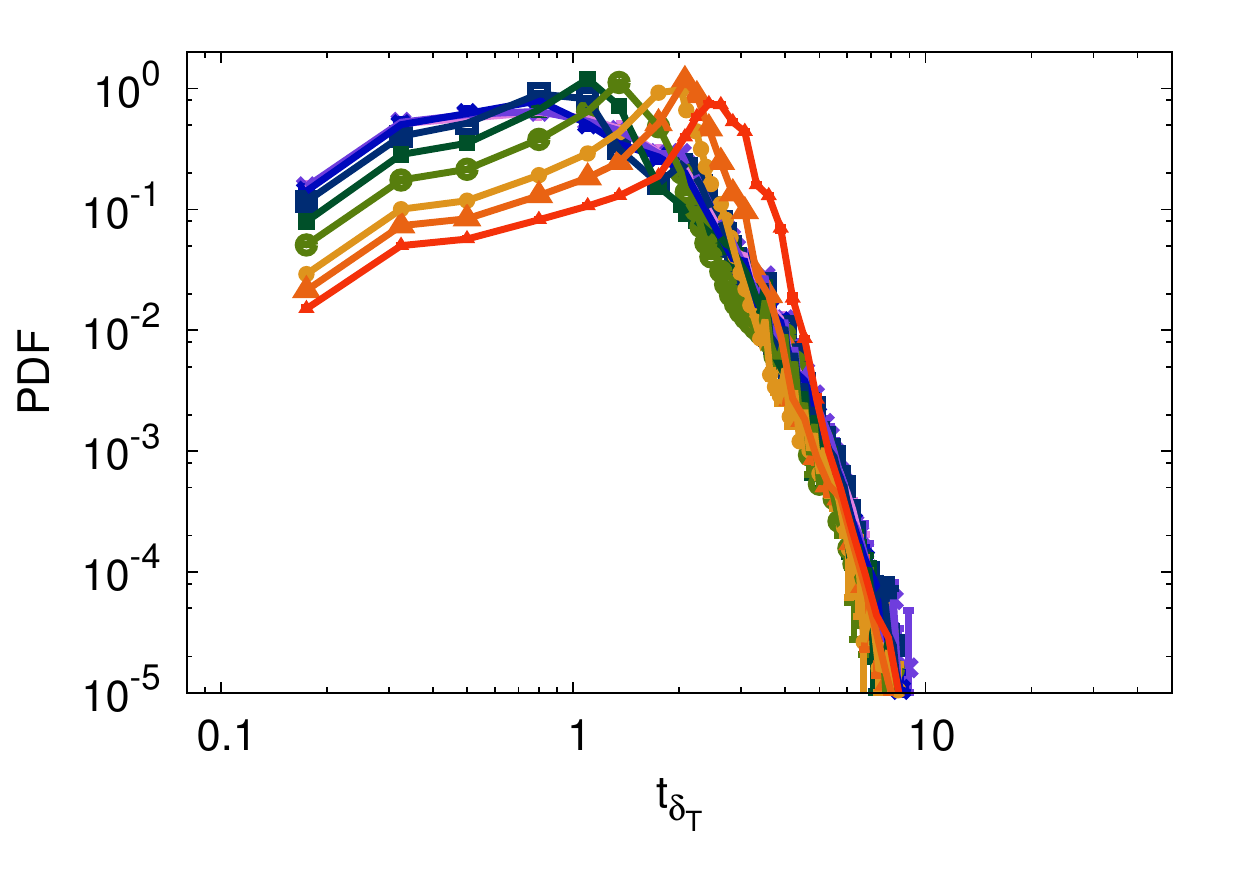} \label{fig:timepdfrun3} }

\caption{PDFs of the particle residence time, $t_{\delta_T}$, inside the thermal boundary layer (BL) at the plates for different values of $K$: (a) $K = 1.1$, (b) $K = 2$ and (c) $K = 10$ and various $\tau_T$ as reported in the legend of panel (a) (see also table \ref{tab:particleprops}). The thermal BLs have a thickness of $\delta_T = 0.022 H$. Errors are estimated as the deviation between the PDFs measured in the thermal BL at the top and bottom plates. Error bars fall within the symbol size.}
\label{fig:pdfstime}
\end{figure}

The thermal response time, $\tau_T$, not only influences the temperature difference between particles and the surrounding fluid, but also the time particles reside at the plates before they will escape from the BL due to the buoyancy force. To understand this relation, statistics of the residence time of particles inside the thermal BLs, $t_{\delta_T}$, are computed for different values of $\tau_T$ and $K$, where the resulting PDFs are shown in fig. \ref{fig:pdfstime}. For $\tau_T \gtrsim 1$, the PDFs display a clear peak suggesting that there is a well-defined characteristic time that particles spend inside the thermal BLs. This peak shifts to the right for increasing $\tau_T$, so this characteristic residence time increases with increasing $\tau_T$ as expected. For $\tau_T \lesssim 1$, the PDFs overlap indicating that here $t_{\delta_T}$ is largely independent of $\tau_T$. When comparing the different values of $K$, it is observed that smaller values of $t_{\delta_T}$ are  measured for larger values of $K$,  due to particles with a larger thermal expansion coefficient having a quantitatively larger response on temperature fluctuations in the fluid in terms of their mass density.   \par
\begin{figure}
\includegraphics[width=0.5\textwidth]{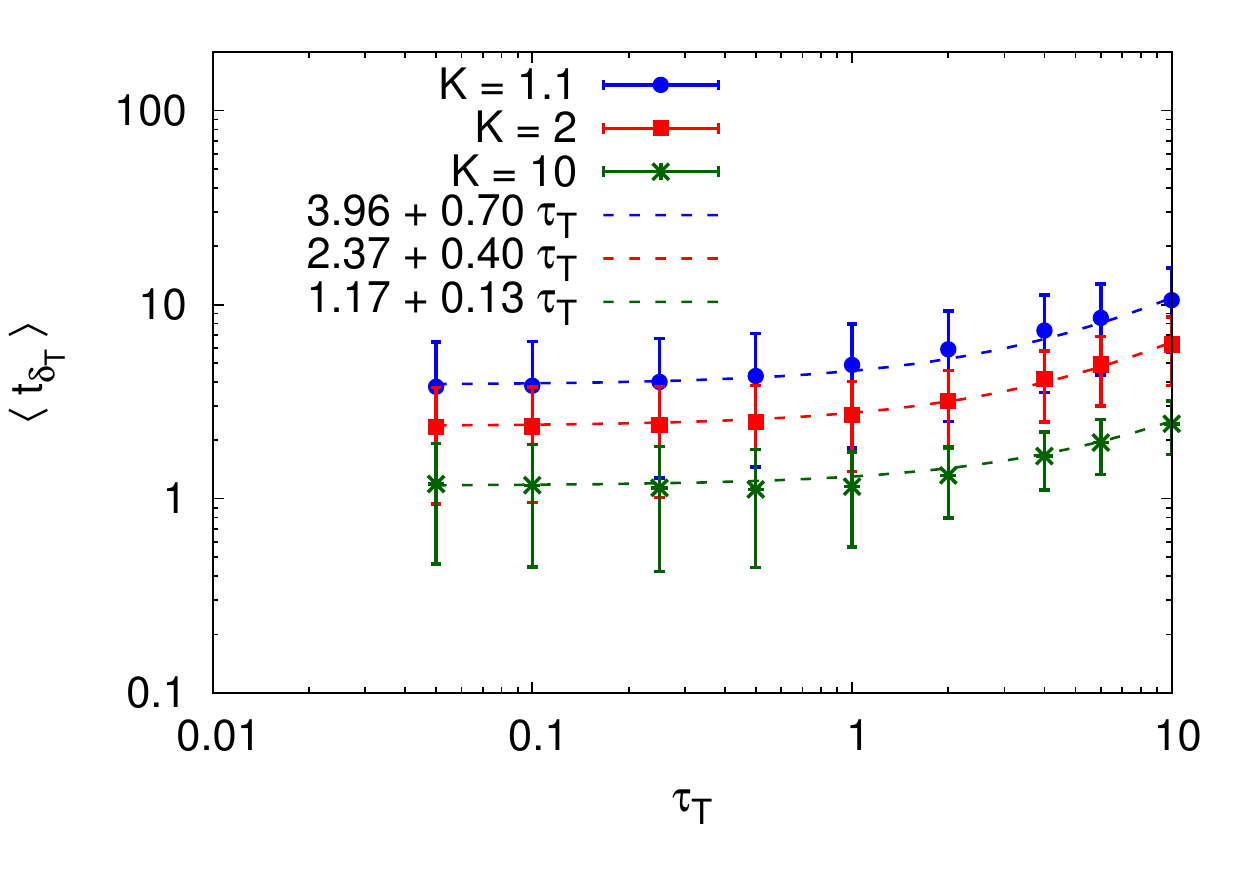}

\caption{The average residence time, $\langle t_{\delta_T} \rangle$ of thermally responsive particles in the thermal boundary layers at the plates as a function of $\tau_T$, for $K = 1.1$ (blue), $K = 2$ (red) and $K = 10$ (green) for DNS (symbols). The dashed lines show fits of the function $y = a(K) + b(K) x$, where the fitting coefficients $a(K)$ and $b(K)$ are as reported in the legend.}
\label{fig:tdvstaufit}
\end{figure}

From the PDFs in fig. \ref{fig:pdfstime} it is expected that $t_{\delta_T}$ depends strongly on $\tau_T$ and $K$. In fig. \ref{fig:tdvstau} we show the average residence time of particles inside the thermal BLs at the horizontal plate, $\langle t_{\delta_T} \rangle$, as a function of $\tau_T$ and for different values of $K$.  Each individual particle can cross the BL multiple times and the average is therefore taken over the total number of times all particles accumulatively cross the BLs. Two regimes can be distinguished in fig. \ref{fig:tdvstaufit}, where for $\tau_T \lesssim 1$ the thermal BL residence time is constant, for $\tau_T \gtrsim 1$ the values are increasing with increasing $\tau_T$. Also, when comparing the three different values of $K$, the residence time is found to decrease with increasing $K$. These trends are consistent with the PDFs shown in fig. \ref{fig:pdfstime} and confirm that  the number of particles inside the BL in fig. \ref{fig:dens} is indeed directly related to the time particles spend inside the thermal BL. \par 

Based on the transition from a constant to a ballistic regime, we can estimate the thermal BL residence time as 
\begin{equation}
t_{\delta_T} = a(K) + b(K) \tau_T, \label{eq:tfit}
\end{equation}
where both $a(K)$ and $b(K)$ are coefficients depending on $K$. In the limit of small thermal response times, $\tau_T \rightarrow 0$, this equation becomes $t_{\delta_T} = a(K)$ and thus a constant depending only on $K$. When $\tau_T \rightarrow \infty$ a ballistic behavior $t_{\delta_T}   = b(K)\tau_T$ is found. We perform a fit based on eq. \eqref{eq:tfit} on the DNS data as shown by the dashed lines in fig. \ref{fig:tdvstaufit} and find that the thermal BL residence time indeed depends on $\tau_T$ as in eq. \eqref{eq:tfit}. \par
\subsection{Simple 1-dimensional model}
To understand in more detail how the dynamics of thermally responsive particles depends on $\tau_T$ and $K$, we develop a simple 1-dimensional (1D) model for the thermally responsive particles. The thermal response time, $\tau_T$, influences the temperature of particles through the thermal inertia (eq. \eqref{eq:TEparticles} for the DNS), while the parameter $K$ determines the density ratio $\widetilde{\beta}$, which is determining the buoyancy force in eq. \eqref{eq:MRparticles}. Therefore, we develop a 1D model for their vertical position in which particles experience both thermal inertia and a buoyancy force: 
\begin{align}
\frac{ \text{d} z}{\text{d} t} &= w(t),  \label{eq:1dz} \\[0.4cm] 
\frac{ \text{d} w(t)}{\text{d} t} &= -g \left( 1 - \frac{1}{\widetilde{\beta}(t)} \right), \quad \widetilde{\beta} = \frac{1 - \alpha_p(T_p(t) - 0.5)}{1 - \alpha_f(T_f - 0.5)} \label{eq:1dw} \\[0.4cm]
\frac{ \text{d} T_p(t)}{\text{d} t} &= \frac{1}{\tau_T}  \left( T_f - T_p(t)\right),  \label{eq:1dTp}
\end{align}
where $w$ is the velocity of particles and we use that $\langle T_p \rangle = \langle T_f \rangle = T_m = 0.5$. The velocity, $w$, is set to zero at $z=0$ and $z=1$. The fluid temperature $T_f$ is now an input parameter of this simple 1D model. In RBC the temperature profile typically shows a large temperature gradient in the thermal BLs, while the temperature in the bulk equals the average temperature \cite{Kerr1996,Ahlers2009}. Therefore we prescribe a constant mean temperature profile with linear temperature gradients inside the thermal BLs and a constant temperature of $T_f = 0.5$ in the bulk:
\begin{align*}
T_f(z)=\begin{cases}
               1 - 0.5 z /\delta_T, & z \leq \delta_T\\
               0.5, & \delta_T < z < 1-\delta_T\\
               0.5(1-z)/\delta_T & z \geq 1 - \delta_T
            \end{cases}
\end{align*}
The thermal BL thickness is set to $\delta_T = 0.022H$, equal to the thermal BL thickness measured from the DNS. \par

\begin{figure*}
\subfloat[]{\includegraphics[width=0.5\textwidth]{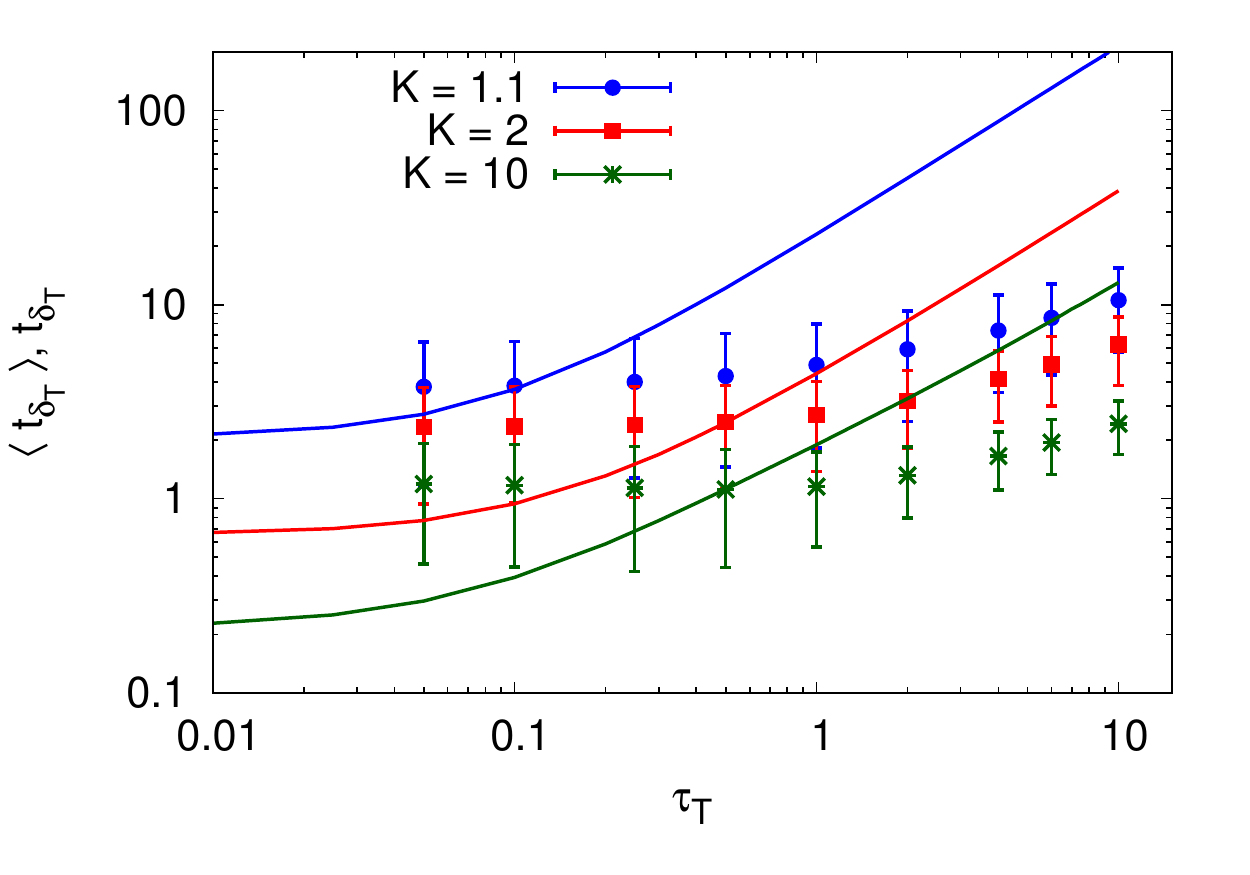}  \label{fig:tdvstau}}
\subfloat[]{\includegraphics[width=0.5\textwidth]{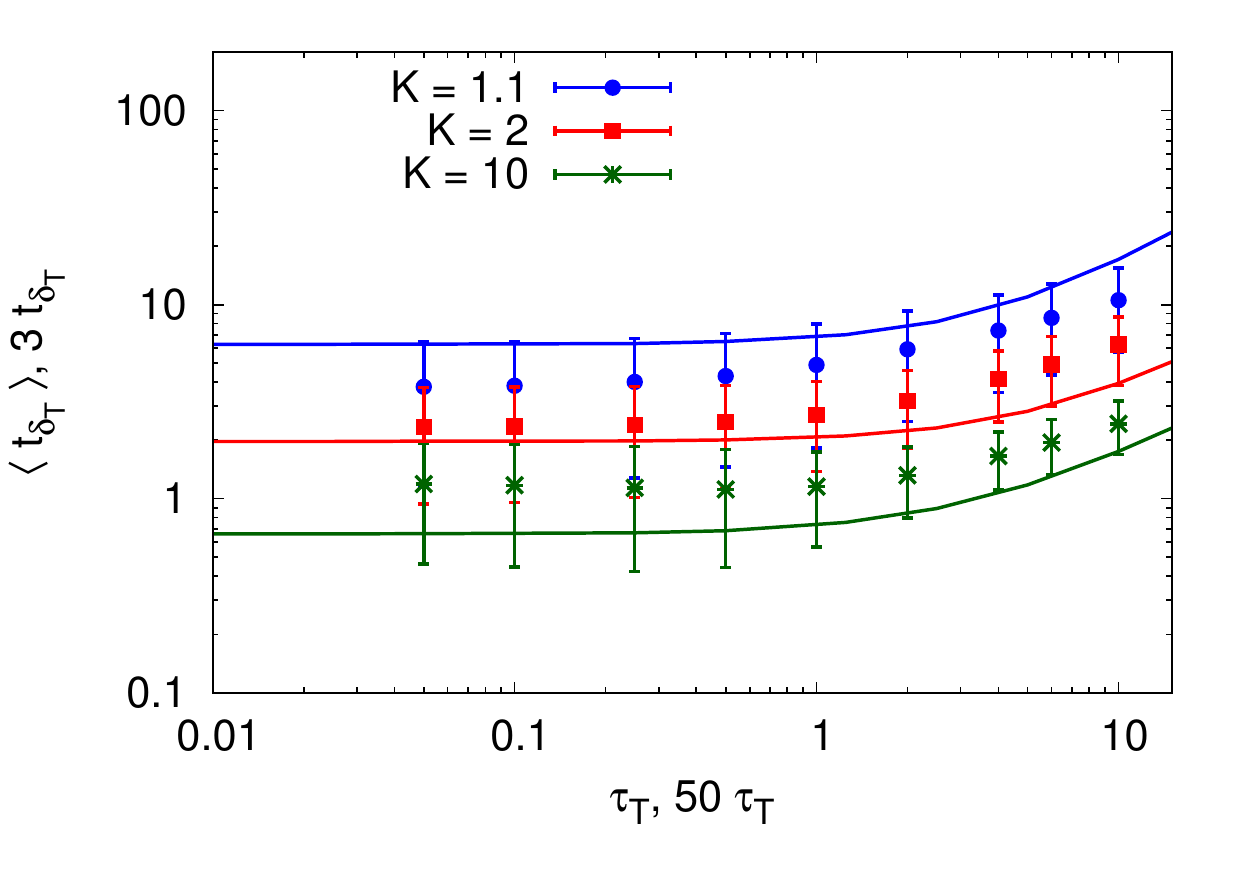}  \label{fig:tdvstaunorm}}

\caption{(a) The average residence time, $\langle t_{\delta_T} \rangle$, for the DNS (symbols) and the residence time, $t_{\delta_T}$, for the 1D model (lines) of thermally responsive particles in the thermal boundary layers at the plates as a function of $\tau_T$, for $K = 1.1$ (blue), $K = 2$ (red) and $K = 10$ (green).  (b) The same data, but now the axes are re-scaled for the 1D model data where the vertical axis is multiplied by 3 and the horizontal axis is multiplied by 50. }
\end{figure*}

From the 1D model, we compute the residence time inside the BLs in the same parameter range as in the DNS, $0.05 < \tau_T < 10$ and $K = \{1.1; 2; 10\}$, by numerically integrating eqs. \eqref{eq:1dz}--\eqref{eq:1dTp} using a second order Adams--Bashforth scheme. Note that the model gives one unique solution and therefore the output is given in terms of $t_{\delta_T}$ and not as an ensemble average $\langle t_{\delta_T} \rangle$ as in the DNS. The model results (lines), together with the DNS data (symbols), are shown in fig. \ref{fig:tdvstau}. We observe that the model captures the trend of decreasing $t_{\delta_T}$ with increasing $K$ as observed in the DNS. Also the trend with $\tau_T$ is recovered where $t_{\delta_T}$ is constant for smaller $\tau_T$ and is monotonically increasing with $\tau_T$ for larger $\tau_T$. \par 
Let us discuss these two regimes in more detail by looking at the behavior of the model in the limit of (i) small thermal response times, $\tau_T \rightarrow 0$, and (ii) large thermal response times, $\tau_T \rightarrow \infty$: 
\begin{enumerate}[i]
  \item $\tau_T \rightarrow 0$: When the thermal response time is zero, particles are instantaneously adapting their temperature to that of the surrounding fluid such that always $T_p = T_f$. Substituting this into eq. \eqref{eq:1dw} gives us 
 \begin{equation}
a = \frac{\text{d} w(t)}{\text{d} t} = -g \left( 1 - \frac{1-\alpha_f(T_f - 0.5)}{1 - \alpha_p(T_f - 0.5)} \right). \label{eq:asmalltaut}
 \end{equation}
Given that $\alpha_p > \alpha_f$ and $\alpha_p T_p' < 1$, we find that in the  BL at the top plate where $T_f > 0.5$ the acceleration is negative ($a < 0$) while in the bottom BL where $T_f > 0.5$ it is positive ($a > 0$).   Equation \eqref{eq:asmalltaut}  does not depend on $\tau_T$ and consequently also the residence time inside the BL (for $\tau_T \rightarrow 0$) is expected to be independent of $\tau_T$ and to only depend on the thermal expansion coefficient and thus on $K$.  This is exactly what we found in fig. \ref{fig:tdvstau} for both the 1D model and the DNS in the limit of small $\tau_T$.
  \item $\tau_T \rightarrow \infty$: For very large thermal response times eq. \eqref{eq:1dTp} becomes $\text{d} T_p / \text{d} t = 0$ and the temperature of particles will be constant and independent of time, such that $T_p = T_p(t=0)$. The initial temperature condition is thus fully determining the particle temperature. A particle initially positioned in the bulk will start to move upwards or downwards depending on its initial  temperature condition, $T_p(t=0)$. It can be computed that when a particle initially moves upwards, the acceleration and velocity in the top BL are positive such that a particle will end up in the top BL and will get stuck there. Oppositely,  an initially downwards moving particle will experience a downwards acceleration and velocity in the bottom BL and will get stuck inside the bottom BL. This means that in the limit of $\tau_T \rightarrow \infty$, $t_{\delta_T} \rightarrow \infty$.  
\end{enumerate}
These limits are consistent with eq. \eqref{eq:tfit}, that was shown to capture the trend of the DNS data. Although not shown here, we verified that the same fitting procedure works for the 1D model confirming that also $t_{\delta_T}$ computed from the 1D model follows eq. \eqref{eq:tfit}. \par 

So, both in the model and in the DNS we observe a transition from a constant residence time to a ballistic regime, where the residence time is increasing with increasing thermal response time. However, the transition between the constant and ballistic regimes, occurs at a different value of $\tau_T \approx 0.1$ in the 1D model, compared to the DNS (where the transition occurs around $\tau_T \approx 1$) in fig. \ref{fig:tdvstau}. This might be related to the model being in 1D, while in the DNS particles move in a 3D flow field. As a result the time scales might not be one-to-one comparable. Moreover in the DNS particles are additionally transported towards and away from the plates by the LSC, an effect that is not included in the simple 1D model. Since we do not expect the model and DNS data to match one-to-one, we can re-scale the vertical and horizontal axis for the model results in fig. \ref{fig:tdvstau}. The DNS data, together with the re-scaled data of the 1D model, are shown in fig. \ref{fig:tdvstaunorm} and now the model matches the DNS results within the error bars. All together we argue that this, though very simple 1D model, captures the trends found in the DNS surprisingly well. 
%
%%%%%%%%%%%%%%%%%%%%%%%%% CONCLUSION  %%%%%%%%%%%%%%%%%%%%%%%%%%%%%
\section{Conclusions} \label{sec:Conclusion}
We have studied the dynamics of thermally responsive particles in Rayleigh-B\'{e}nard convection. Particles are experiencing both mechanical and thermal inertia, and both fluid and particles exhibit thermal expansion where the thermal expansion coefficient of particles is larger than that of the fluid.  Now, particles near the hot bottom plate become lighter than the fluid and particles at the top plate become heavier than the fluid.  It is verified that this induces a motion away from the plates, resulting in particles re-suspending from the BLs into the bulk. \par

This dynamics results in a non-homogeneous distribution of particles throughout the RBC cell. In particular, a regime of thermal response times and thermal expansion coefficients is found where the number of particles at the plate is enhanced compared to a uniform distribution. We have shown that this enhancement is already reached for an increase in the thermal expansion coefficients of particles compared to that of the fluid  of ten per cent; $K = \alpha_p / \alpha_f =  1.1$. This ratio of thermal expansion coefficients can be achieved in realistic systems, for example gel-like particles in water or oil--water systems.  \par 

Upon increasing $K$, the number of particles at the plates is decreasing, since the particle density responds much stronger to the temperature fluctuations. A regime of large $K$ and small $\tau_T$ is found where particles escape the BLs almost immediately and where the number of particles inside the thermal BLs is even lower than the uniform distribution. This depletion in the BLs leads to an enhanced number of particles inside the bulk. Increasing $\tau_T$ has an opposite effect; particles need more time to warm up (cool down) at the bottom (top) plate, increasing the number of particles at the plates and decreasing the number of particles in the bulk. \par

The number of particles at the plates is expected to depend on the time particles spend inside the thermal BLs at the plates. By quantifying this residence time, $t_{\delta_T}$, it has been shown that particles do spend a characteristic time inside these BLs, that is moreover depending on $\tau_T$ and $K$. In particular, the ensemble average $\langle t_{\delta_T} \rangle$ is increasing with decreasing $K$.  For all values of $K$, $\langle t_{\delta_T} \rangle$ is constant for $\tau_T \lesssim 1$ and is increasing with increasing $\tau_T$ for $\tau_T \gtrsim 1$ in the DNS. This trend is confirmed when performing a fit of the function $y = a(K)  + b(K)  x$ on the DNS data for each value of $K$. A simple 1D model is developed, where the motion of thermally inertial particles depends exclusively on the buoyancy force, and again both particles and fluid exhibit thermal expansion with $\alpha_p > \alpha_f$. This model is shown to capture the trends very well; again the thermal BL residence time is constant for smaller $\tau_T$ and increasing with increasing $\tau_T$ for larger $\tau_T$, only now the transition occurs at a smaller $\tau_T \approx 0.1$. Also the shift of the curves to lower values of $t_{\delta_T}$ for larger values of $K$ is captured well by the model. When re-scaling the data of the model the DNS and model results match within the error bars, confirming that the model captures the observed trends well. The simple 1D model can thus be used to better understand the interplay between thermal inertia and the buoyancy-driven vertical motion of particles. \par 

We have studied how thermal inertia influences the dynamics of thermally responsive particles, using a point-particle approach. The dynamics in this point-particle model is already rich and there are many parameters involved. In nature, however, multi-phase fluid systems with different thermal properties for the different phases can become even more complex; \textit{e.g.} phase transitions in convection in the core of the earth or the presence of deformable vapor bubbles in boiling convection. To study these highly complex systems more advanced numerical techniques, with much higher numerical costs, are necessary. Here we have however shown that DNS with a point-particle approach is able to give insight into the influence of thermal inertia on the distribution and the temperature statistics  of inertial particles in a thermally driven flow where the dispersed phase has different thermal properties than  the carrier fluid.
%
%%%%%%%%%%%%%%%%%%%%%%%%% ACKNOWLEDGEMENTS %%%%%%%%%%%%%%%%%%%%%%%%
\section*{Acknowledgments}
This work is supported by  the Nederlandse Organisatie voor Wetenschappelijk Onderzoek I (NWO-I), the Netherlands. The authors gratefully acknowledge the support of NWO for the use of supercomputer facilities (Cartesius) under Grant No. 16289. EU-COST action MP1305 `Flowing matter' is kindly acknowledged. 
%
%%%%%%%%%%%%%%%%%%AUTHOR CONTRIBUTIONS %%%%%%%%%%%%%%%%%%%%%%%%
%
\section*{Author contribution statement}
All authors contributed to the design of the research project. KA performed the numerical simulations and analyzed the data and all authors discussed the results. KA wrote the first draft which is modified with the support and feedback of RK, FT and HC. All the authors have read and approved the final manuscript.
%
%%%%%%%%%%%%%%%%%%%%%%%%% BIBLIOGRAPHY  %%%%%%%%%%%%%%%%%%%%%%%%%%%
%
\FloatBarrier
\bibliographystyle{plain}
\bibliography{mybib}{}
%\begin{thebibliography}{41}%
%\end{thebibliography}%
%
\end{document}